\documentclass[12pt,draftclsnofoot,onecolumn]{IEEEtran}

\usepackage{amsmath,graphics,amssymb,epsfig,subfigure,color,amsthm,cite}
\usepackage{array}
\usepackage{multirow}
\usepackage{enumerate}
\usepackage{algorithmic}
\usepackage{algorithm}
\usepackage{bm}
\usepackage{float}
\usepackage{makecell}

\newtheorem{prop}{Proposition}

\newcommand{\argmax}{\operatornamewithlimits{argmax}}

\makeatletter
\newcommand{\vast}{\bBigg@{4.5}}
\newcommand{\Vast}{\bBigg@{7.5}}
\makeatother

\begin{document}
	\title{A Compressive Sensing Approach for Federated Learning  over 
	Massive MIMO Communication Systems}

	\author{Yo-Seb Jeon, Mohammad Mohammadi Amiri, Jun Li, and H. Vincent Poor
		\thanks{Y.-S. Jeon is with the Department of Electrical Engineering, POSTECH, Pohang, Gyeongbuk 37673, South Korea (e-mail: yoseb.jeon@postech.ac.kr).}
		\thanks{M. M. Amiri and H. V. Poor are with the Department of Electrical Engineering, Princeton University, Princeton, NJ 08544 (e-mails: mamiri@princeton.edu, poor@princeton.edu).}
		\thanks{J. Li is with School of Electronic and Optical Engineering, Nanjing University of Science and Technology, Nanjing 210094, China (e-mail: jun.li@njust.edu.cn).}
	}
	\vspace{-2mm}	
	
	\maketitle
	\vspace{-12mm}

	\begin{abstract} 
		Federated learning is a privacy-preserving approach to train a global model at a central server by collaborating with wireless devices, each with its own local training data set. In this paper, we present a compressive sensing approach for federated learning over massive multiple-input multiple-output communication systems in which the central server equipped with a massive antenna array communicates with the wireless devices. One major challenge in system design is to reconstruct local gradient vectors accurately at the central server, which are computed-and-sent from the wireless devices. To overcome this challenge, we first establish a transmission strategy to construct sparse transmitted signals from the local gradient vectors at the devices. We then propose a compressive sensing algorithm enabling the server to iteratively find the linear minimum-mean-square-error (LMMSE) estimate of the transmitted signal by exploiting its sparsity. We also derive an analytical threshold for the residual error at each iteration, to design the stopping criterion of the proposed algorithm. We show that for a sparse transmitted signal, the proposed algorithm requires less computationally complexity than LMMSE. Simulation results demonstrate that the presented approach outperforms conventional linear beamforming approaches and reduces the performance gap between federated learning and centralized learning with perfect reconstruction.
	\end{abstract}

	
	\section{Introduction}\label{Sec:Intro}
	Machine learning has attracted significant interest as a breakthrough for emerging applications of wireless communications \cite{Jiang:17,Zhu:20,He:19,Jeon:20,Oshea:17,Jeon:18}. 
	The fundamental idea of machine learning for wireless communications is to learn a model (e.g., input-output relation) based on large amounts of data and computing power. It has been demonstrated that the learned model can be exploited either to improve the performance of conventional model-based techniques (e.g. \cite{He:19,Jeon:20}) or to describe an unknown input-output relation whose characterization was otherwise challenging due to mathematical intractability (e.g., \cite{Oshea:17,Jeon:18}). The most common and widely adopted form of machine learning is \textit{centralized learning} in which a central server equipped with sufficient storage and computing power has full access to the entire data set. Unfortunately, centralized learning is infeasible in many applications of wireless communications. The major reason is that data sets are usually generated at wireless devices, but transmitting them to the central server is highly limited by both the amount of radio resources and communication latency allowed in the applications. 
	Particularly, this problem becomes more severe as the data size and the model complexity increase. In addition, sending the data may not be allowed in privacy-sensitive applications such as social networking, e-health, and financial services. 

	Recently, federated learning has drawn increasing attention as a viable solution to overcome the limitations of centralized learning \cite{Konecny:15,Konecny:17,McMahan:17,Lin:18}, in which a global model at a central server is collaboratively trained by multiple devices each with its own local data set. 
	The major advantage of federated learning is a significant reduction in communication overhead as the devices only send an updated model instead of the whole data set. In addition, this approach preserves the privacy of the devices because the data is kept where it is generated while a central server has no direct access to the local data sets. Distributed machine learning also provides similar advantages, but federated learning focuses on a more practical setting which may include unbalanced and non-identically-distributed data sets, unreliable communication, and massively distributed data \cite{Konecny:15}. Thanks to the advantages and the practicality of federated learning, it has been adopted as a key enabler for emerging applications of wireless communications \cite{Niknam:19,Samarakoon:19,Chen:19}. For example, in \cite{Samarakoon:19}, federated learning is applied to learn the statistical properties of vehicular users in wireless vehicular networks. Another example is introduced in \cite{Chen:19} which applies federated learning to learn the locations and orientations of the users in wireless virtual reality networks.

	There also exist recent studies that seek to enable and optimize federated learning over wireless communication systems by considering the physical characteristics of wireless channels
	\cite{Nishio:19,Liu:19,Howard:19,Howard:20,Amiri:arXiv:20,Chen:19-2,Amiri:ISIT,Amiri:SPAWC,Zhu:TWC,Wei:20}. 
	A device scheduling problem is studied in \cite{Nishio:19,Liu:19,Howard:19,Howard:20,Amiri:arXiv:20} based on various scheduling criteria, while a joint resource allocation and user scheduling problem is tackled in \cite{Chen:19-2}. Transmission and reception techniques for federated learning are developed for a simple Gaussian multiple access channel (MAC) in \cite{Amiri:ISIT} and for a fading MAC in \cite{Amiri:SPAWC}. A transmission technique for the fading MAC is also proposed in \cite{Zhu:TWC} jointly with a device scheduling method. 
    	To improve the robustness against channel fading and noise effects in wireless environments, the use of multiple antennas at a central server and/or wireless devices is considered in \cite{Amiri:GlobalSIP,Vu:19,Yang:19,Wen:19}. 
    	The effect of multiple antennas on the performance of federated learning is investigated in \cite{Amiri:GlobalSIP} under the assumption that receive beamforming is employed by a multi-antenna central server. 
    	In this work, the use of the receive beamforming is shown to be an effective approach when the server is equipped with a sufficiently large number of antennas.  
    	Under the same assumption as in \cite{Amiri:GlobalSIP}, training time optimization for federated learning is studied in \cite{Vu:19}.
    	Transmit and/or receive beamforming for over-the-air computation are considered in \cite{Yang:19,Wen:19} which can be utilized to reduce latency in federated learning. 
    	Specifically, joint optimization for device selection and receive beamforming is proposed in \cite{Yang:19}, while joint design of transmit and receive beamforming is proposed in \cite{Wen:19}.
    	The common limitation of the existing works in \cite{Amiri:GlobalSIP,Vu:19,Yang:19,Wen:19}, however, is that they only consider linear beamforming approaches to design the reception technique at the multi-antenna server.
    	Unfortunately, such linear beamforming approach is not optimal in terms of the estimation performance at the server in general; thereby, further investigation on the reception technique is still needed to maximize the estimation performance at the multi-antenna central server in federated learning over wireless communication systems. 


    In this work, we study federated learning over massive multiple-input multiple-output (MIMO) communication systems, in which a global model is trained through collaboration of multiple wireless devices, each with its own local training data set, with a central server equipped with a massive number of antennas. 
    	One of the challenges in system design is to reconstruct local gradient vectors accurately at the central server, which are computed-and-sent from the wireless devices.
    	Our key observation to overcome this challenge is that the local gradient vectors are likely to be sparse in the considered federated learning framework. 
    	Motivated by this observation, we present a compressive sensing approach that exploits the sparsity of the local gradient vectors, to enable efficient reconstruction of them at the central server.
    	To the best of our knowledge, this work is the first attempt to establish a compressive sensing approach for a multi-antenna central server in federated learning, to exploit the sparsity in a spatial-device domain. 
        Note that the existing works in \cite{Amiri:ISIT,Amiri:SPAWC} also consider a compressive sensing approach, but for a single-antenna server, while exploiting the sparsity in a different domain.
	The major contributions of this paper are summarized as follows:
	\begin{itemize}
	    \item We establish a transmission strategy to construct sparse transmitted signals from the local gradient vectors at the wireless devices. 
		The basic idea of this strategy is to permute the local gradient vectors using different patterns across the wireless devices. 
		The major advantage of our transmission strategy is that when the local gradient vectors are sparse, only a small subset of devices simultaneously transmit non-zero gradient elements at each radio resource, which results in a sparse transmitted signal. 
		Using simulations, we demonstrate that the use of our transmission strategy significantly improves the sparsity of the transmitted signal in the considered federated learning framework. 
		\item We propose a compressive sensing algorithm enabling the central server to efficiently estimate the transmitted signal by exploiting its sparsity.
		The basic idea of the proposed algorithm is to iteratively find the linear minimum-mean-square-error (LMMSE) estimate of non-zero elements of the transmitted signal.
		In the proposed algorithm, to properly determine the LMMSE estimate at the server, we establish a statistical model for the transmitted signal based on a large-scale approximation and a statistical feature obtained from our transmission strategy.
		We also derive an analytical threshold for the residual error at each iteration, to design the stopping criterion of the proposed algorithm.
		Simulation results demonstrate that the proposed compressive sensing algorithm with our transmission strategy efficiently reduces the performance gap between federated learning and centralized learning with perfect reconstruction, when the size of the mini-batch employed at each device is relatively small.
		\item We compare our compressive sensing approach with linear beamforming approaches that can be employed to reconstruct local gradient vectors at the multi-antenna central server. To this end, we introduce proper modifications of the conventional maximum ratio combining (MRC) and LMMSE methods and present their limitations for the use in federated learning over a massive MIMO system. We then compare the computational complexity of the proposed algorithm with those of the linear beamforming methods. Our key finding is that the proposed algorithm requires less complexity than the LMMSE method when the transmitted signal is sparse, which is verified through both analytical and numerical results. We also demonstrate that the proposed algorithm outperforms linear beamforming methods in terms of the classification accuracy of federated learning, using simulations.
	\end{itemize}

	\subsubsection*{Notation}
	Upper-case and lower-case boldface letters denote matrices and column vectors, respectively.
	$\mathbb{E}[\cdot]$ is the statistical expectation,
	$\mathbb{P}(\cdot)$ is the probability,
	$(\cdot)^{\sf T}$ is the transpose,
	$(\cdot)^{\sf H}$ is the conjugate transpose,
	$\lceil \cdot \rceil$ is the ceiling function,
	$\lfloor \cdot \rfloor$ is the floor function,
	and $|\cdot|$ is the absolute value.
	${\sf Re}\{\cdot\}$ and ${\sf Im}\{\cdot\}$ denote real and imaginary components, respectively.
	$|\mathcal{A}|$ is the cardinality of set $\mathcal{A}$.
	$({\bf a})_i$ represents the $i$-th element of vector ${\bf a}$.
	$\|{\bf a}\|\!=\!\sqrt{{\bf a}{\bf a}^{\sf H}}$ is the Euclidean norm of vector ${\bf a}$.
	$\mathcal{CN}({\bm \mu},{\bf R})$ represents the distribution of a circularly symmetric complex Gaussian random vector with mean vector ${\bm \mu}$ and covariance matrix ${\bf R}$.
	${\bf 0}_n$ and ${\bf 1}_n$ are an $n$-dimensional vectors whose elements are zero and one, respectively.
	$\mathbb{R}$ is the set of real numbers.

	\begin{figure*}[t]
		\centering 
		{\epsfig{file=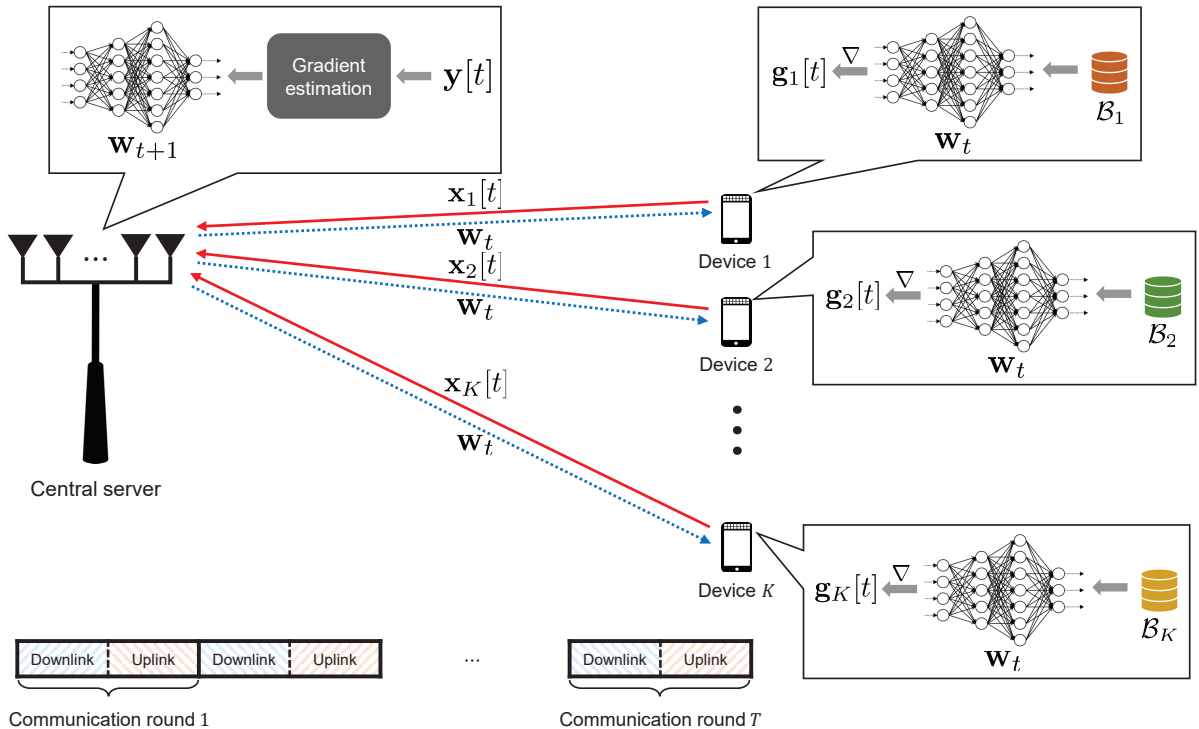, width=16cm}}
		\caption{An illustration of federated learning over a TDD massive MIMO communication system.} \vspace{-7mm}
		\label{fig:System}
	\end{figure*}

	\section{System Model}
	We consider federated learning over a time-division-duplex (TDD) massive MIMO communication system in which a central server equipped with $M$ antennas trains a global model by collaborating with $K$ single-antenna wireless devices, as illustrated in Fig. \ref{fig:System}. 
	In this system, the server and the wireless devices share a global model (e.g., neural network) represented by a parameter vector ${\bf w}\in{\mathbb R}^{N_{\rm w}}$, but only the devices have training data samples to train the global model. We denote $\mathcal{B}_k$ as a local training data set available at device $k$ which consists of $|\mathcal{B}_k|$ training data samples for $k\in\mathcal{K}=\{1,\ldots,K\}$. We model the wireless channel from the devices to the server by an $L$-tap channel impulse response (CIR).
    	We assume that this CIR is perfectly known at the server via uplink channel training\footnote{To enable accurate uplink CSI at the server, each device may need to send more than $K$ pilot signals, but this overhead is still negligible compared to the overhead required for gradient vector transmission since typically $N_{\rm w}\gg K$ in the federated learning framework.} performed once per each uplink transmission and remains constant until the next downlink transmission. Note that this is a common assumption in TDD massive MIMO systems \cite{mMIMO1,mMIMO2}.    	
	We adopt an orthogonal frequency division multiplexing (OFDM) modulation with $N_{\rm sub}$ subcarriers to obtain parallel subchannels without inter-symbol interference.

	We assume that the global model is trained using a gradient-based iterative algorithm (e.g., stochastic gradient descent algorithm or Adam optimizer \cite{ADAM}). Each iteration of the algorithm corresponds to one communication round that consists of uplink and downlink phases, as illustrated in Fig. \ref{fig:System}. Let $T$ be the number of iterations and ${\bf w}_t$ be the parameter vector at iteration $t$ of the algorithm. 
	During the uplink phase at communication round $t$, the server broadcasts ${\bf w}_t$ to the wireless devices. Then during the uplink phase, the wireless devices transmit their local gradient vectors to the server.	
	In this work, we focus only on a transmission/reception strategy for the uplink phase, while assuming that the broadcasting of ${\bf w}_{t}$ in the downlink phase is error free, as assumed in most literature \cite{Amiri:ISIT,Amiri:SPAWC,Zhu:TWC,Yang:19,Amiri:GlobalSIP}. Under this assumption, all devices have a globally consistent parameter vector ${\bf w}_t$ for all $t\in\mathcal{T}=\{1,\ldots,T\}$.


	We present the transmission procedure of each wireless device during the uplink phase. 
	    In this work, we assume that the local gradient vector at each device is computed over only a small fraction of its local data set, considering a limited computing power and stringent latency constraint at the wireless devices in practical communication systems. 
	Let $\mathcal{B}_k[t] \subset \mathcal{B}_k$ be a set of the samples selected by device $k$ to compute the local gradient vector at communication round $t$. Then a local gradient vector at device $k$ is computed as 
	\begin{align}\label{eq:local_grad}
	{\bf g}_k[t] = \frac{1}{|\mathcal{B}_k[t]|} \sum_{b\in\mathcal{B}_k[t]} \nabla f({\bf w}_t,b),
	\end{align}
	where $\nabla f(\cdot,b)$ is the gradient of a loss function computed for the training data sample $b\in\mathcal{B}_k$ defined by the learning task. 
	After computing the local gradient vector, a transmitted signal at device $k$ can be constructed as 
	\begin{align}\label{eq:transmitted_signal}
	    {\bf x}_k[t] = \sqrt{\frac{N_{\rm w}}{\|{\bf g}_k{[t]}\|^2}}{\bf g}_k[t].
	\end{align}
	The scaling operation in \eqref{eq:transmitted_signal} is adopted to ensure that every device has the same transmit power of $\|{\bf x}_k[t]\|^2 = N_{\rm w}$ at each communication round.
        We assume that each uplink phase consists of $\lceil\frac{N_{\rm w}}{N_{\rm sub}}\rceil$ OFDM symbols, each with $N_{\rm sub}$ subcarriers, to support the transmission of $N_{\rm w}$ elements. Under this assumption, the $n$-th element of ${\bf x}_k[t]$, namely $x_k[t,n] \in\mathbb{R}$, is transmitted using the $n$-th radio resource of the uplink phase, corresponding to the $f_n$-th subcarrier of the $u_n$-th OFDM symbol where $f_n =n-(u_n-1)N_{\rm sub}$ and $u_n=\lceil\frac{n}{N_{\rm sub}}\rceil$.

	We now describe the reception procedure of the server during the uplink phase. The received signal associated with the $n$-th radio resource  of the uplink phase  is expressed as 
	\begin{align}
	{\bf y}^{\rm c}[t,n] = \sum_{k=1}^{K} {\bf h}_k^{\rm c}[t,n] x_{k}[t,n] + {\bf z}^{\rm c}[t,n],
	\end{align}
	where ${\bf h}_{k}^{\rm c}[t,n]$ is the channel frequency response vector of device $k$ and ${\bf z}^{\rm c}[t,n] \in \mathcal{C}^{M}$ is the noise signal at the $f_n$-th subcarrier of the $u_n$-th OFDM symbol. We assume that the noise signal at each radio resource is independent and identically distributed (i.i.d.) as $\mathcal{CN}({\bf 0}_{M},\sigma_{\rm c}^2{\bf I}_{M})$. The real-domain equivalent representation of ${\bf y}^{\rm c}[t,n]$ is given by
	\begin{align}\label{eq:received_signal0}
	{\bf y}[t,n] = \sum_{k=1}^{K} {\bf h}_k[t,n] x_k[t,n] + {\bf z}[t,n],
	\end{align}
	where 
	\begin{align*}
	{\bf y}[t,n]&=\big[{\sf Re}({\bf y}^{\rm c}[t,n])^{\sf T},{\sf Im}({\bf y}^{\rm c}[t,n])^{\sf T}\big]^{\sf T}, \\
	{\bf h}_k[t,n]&=\big[{\sf Re}({\bf h}_k^{\rm c}[t,n])^{\sf T},{\sf Im}({\bf h}_k^{\rm c}[t,n])^{\sf T}\big]^{\sf T}, \\
	{\bf z}[t,n]&=\big[{\sf Re}({\bf z}^{\rm c}[t,n])^{\sf T},{\sf Im}({\bf z}^{\rm c}[t,n])^{\sf T}\big]^{\sf T}.
	\end{align*}
	The above representation can be rewritten as 
	\begin{align}\label{eq:received_signal}
	{\bf y}[t,n] = {\bf H}[t,n] {\bf x}[t,n] + {\bf z}[t,n],
	\end{align}
	where
	\begin{align*}
	{\bf H}[t,n]&=\big[ {\bf h}_{1}[t,n],{\bf h}_{2}[t,n],\cdots, {\bf h}_{K}[t,n]\big], \\
	{\bf x}[t,n]&=\big[x_{1}[t,n],x_{2}[t,n],\cdots ,x_{K}[t,n]\big]^{\sf T}.
	\end{align*}
	Note that ${\bf z}[t,n] \sim \mathcal{CN}({\bf 0}_{2M},\sigma^2{\bf I}_{2M})$ where $\sigma^2 = \frac{\sigma_{\rm c}^2}{2}$. Based on the received signals $\{{\bf y}[t,n]\}_{n=1}^{N_{\rm w}}$, the server estimates the transmitted signals $\{{\bf x}[t,n]\}_{n=1}^{N_{\rm w}}$ to obtain the information of the local gradient vectors sent from the wireless devices. A detailed estimation process will be discussed in the sequel. Let $\hat{x}_k[t,n]$ be the estimate of the transmitted signal sent from device $k$ at the $n$-th radio resource. 
	Then by aggregating all the estimates of the transmitted signals, the local gradient vector sent from device $k$ can be reconstructed as
	\begin{align}\label{eq:grad_est}
	\hat{\bf g}_k[t] = \sqrt{\frac{\|{\bf g}_k{(t)}\|^2}{N_{\rm w}}} \hat{\bf x}_k[t],
	\end{align}
	where $\hat{\bf x}_k[t]=\big[\hat{x}_k[t,1],\cdots,\hat{x}_k[t,N_{\rm w}]\big]^{\sf T}$ for $k\in \mathcal{K}$. In \eqref{eq:grad_est}, we assume that the norm of the local gradient vector $\|{\bf g}_k{(t)}\|$ for all $k\in\mathcal{K}$ is known\footnote{
    	To convey the information of the norm of the local gradient vector, each device needs to send one additional real value,
    	    but this overhead is still negligible compared to the overhead required for gradient vector transmission since typically $N_{\rm w}\gg 1$ in the federated learning framework.
	}  at the server. After reconstructing all the local gradient vectors, the server aggregates these vectors to obtain the global gradient vector defined as
	\begin{align}\label{eq:global_grad}
	\bar{\bf g}[t] = \frac{1}{\sum_{j=1}^K |\mathcal{B}_j[t]|}
	\sum_{k=1}^K |\mathcal{B}_k[t]|~\hat{\bf g}_k[t]. 
	\end{align}
	The computing power of each device may not change during the training process, so we assume that $\{|\mathcal{B}_k[t]|\}_{t\in\mathcal{T}}$ is fixed and known at the central server. The global gradient vector in \eqref{eq:global_grad} is utilized to update the parameter vector ${\bf w}_t$. For example, if the central server adopts a gradient descent algorithm, the update of the parameter vector is expressed as
	\begin{align}\label{eq:w_update}
	{\bf w}_{t+1} \leftarrow {\bf w}_t - \eta_t \bar{\bf g}[t],
	\end{align}
	where $\eta_t$ represents the learning rate at iteration $t$.

    	In this paper, we address the mismatch between the local gradient vectors sent from the wireless devices and their estimates reconstructed at the server, i.e., $\hat{\bf g}_k[t]\neq {\bf g}_k[t]$, in federated learning over the massive MIMO systems.
    	The main causes of this mismatch are 1) inter-user interference caused by simultaneous transmission of multiple devices, 2) channel fading naturally occurred in wireless channels, and 3) noise signal in RF chain.
    	This mismatch may harm both the learning accuracy and the convergence rate of federated learning as will be demonstrated in Sec. V. 
    	Therefore, it is essential to reduce this mismatch by developing a proper reception technique at the central server that enables accurate estimation of the local gradient vectors. 



	\section{A Compressive Sensing Approach for \\ Efficient Reconstruction of Local Gradient Vectors}\label{Sec:CS}
    	In this section, we present a compressive sensing approach that allows the central server to efficiently reconstruct the local gradient vectors sent from the wireless devices. To this end, we first discuss the sparsity of the local gradient vectors and then establish a transmission strategy to construct sparse transmitted signals from the local gradient vectors at the wireless devices. Based on this strategy, we propose a compressive sensing algorithm enabling the central server to estimate the transmitted signal by exploiting its sparsity.

	\subsection{Motivation: Sparsity of Local Gradient Vectors}\label{Sec:Sparsity}
	    Our compressive sensing approach is motivated by the sparsity of the local gradient vectors in federated learning over wireless communication systems. 
	    First of all, the average magnitudes of gradient elements are expected to reduce as a training algorithm (e.g., gradient descent algorithm) proceeds; 
	        thereby, the number of zero gradient elements may increase over time.
       	Particularly this number is much larger when employing a ReLU activation function 
	        since the gradient of the ReLU function is zero for any negative-valued input. 
	    We have also observed that the number of zero gradient elements are likely to increase as the size of the mini-batch training data samples utilized to compute the local gradient vector at each device decreases.
	    The rationale behind this observation is that zero gradient elements computed for each training data sample would be associated with a significantly different subset of weights across different samples.
        This observation is particularly relevant for federated learning over wireless communications 
	        since the size of the mini-batch samples at wireless device (e.g., smartphone or IoT device) is expected to be relatively small due to a limited computing power and/or a stringent latency constraint.
        All these observations imply that the local gradient vectors are likely to be sparse in federated learning over practical communication systems.
        It is also noticeable that transmitted signals whose magnitudes are much smaller than the noise level of a communication system can be treated as zero signals 
            since they have negligible impacts on the performance of federated learning.
        Meanwhile, only a few of the largest gradient elements will remain large in the transmission signal at the wireless devices
            since each device normalizes the local gradient vector before its transmission, as can be seen in \eqref{eq:transmitted_signal}. 
        These observations imply that even if the local gradient vectors may not be exactly sparse in certain cases (e.g., during the initial training iterations),
            an approximate sparsity can still be observed at the central server. 

	\begin{figure}[t]
    	\centering 
    	{\epsfig{file=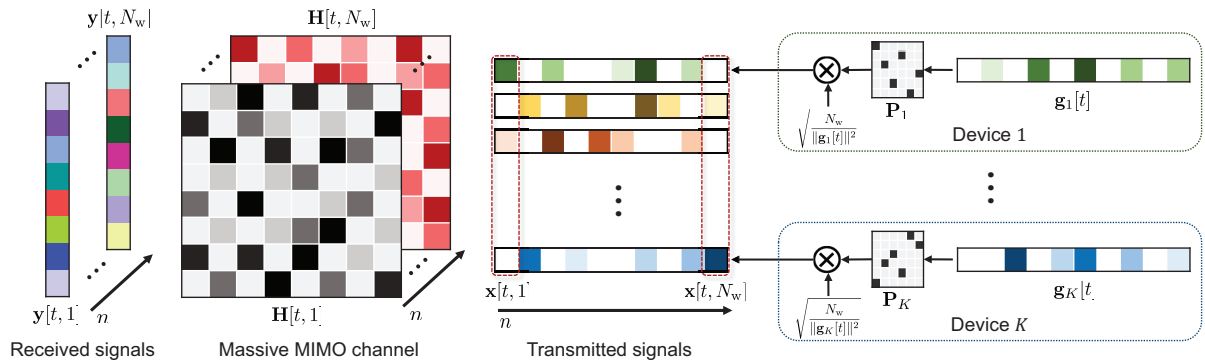, width=16cm}}
    	\caption{An illustration of transmitted signals and the corresponding received signals when employing the random permutation strategy in Sec.~\ref{Sec:Permute}.} \vspace{-3mm}
    	\label{fig:Transmit}
    \end{figure}

	\subsection{Transmission Strategy: Random Permutation}\label{Sec:Permute}
    	Even when the local gradient vectors are sparse, the transmitted signal may not be sparse if a naive transmission procedure (e.g., the procedure introduced in Sec. II) is employed by the wireless devices. 
    	For example, consider an extreme case where only the $n^*$-th element of the local gradient vector is non-zero at all the devices (i.e., ${g}_k[t,n^\star]\neq 0$ for $k\in\mathcal{K}$).
    	In this case, the transmitted signal at the $n^*$-th radio resource, ${\bf x}[t,n^*]$, is not sparse because $x_{k}[t,n^*]$ is non-zero for all $k\in\mathcal{K}$ from \eqref{eq:transmitted_signal}. More generally, if all the local training data sets have an identical distribution, the wireless devices may have similar sparsity patterns \cite{Amiri:ISIT,Amiri:SPAWC}; in this case, the transmitted signals constructed from \eqref{eq:transmitted_signal} may not be sparse.

    	To prevent the loss of the sparsity, we establish a new transmission strategy that allows the wireless devices to construct sparse transmitted signals by preserving the sparsity of the local gradient vectors. The basic idea is to permute the local gradient vectors using different patterns across the wireless devices when constructing the transmitted signal. This strategy is enabled through linear projection using different permutation matrices at different devices. More precisely, the transmitted signal of device $k$ is determined as 
    	\begin{align}\label{eq:transmitted_signal2}
    	    {\bf x}_k[t] = \sqrt{\frac{N_{\rm w}}{\|{\bf g}_k{[t]}\|^2}}{\bf P}_k{\bf g}_k[t],
    	\end{align}
    	where ${\bf P}_k \in \{0,1\}^{N_{\rm w}\times N_{\rm w}}$ is the permutation matrix employed at device $k$ such that ${\bf P}_k {\bf P}_k^{\sf T} = {\bf I}_{N_{\rm w}}$.
    	Then the server reconstructs\footnote{ We assume that the central server has the information of the permutation matrices employed at wireless devices. One possible approach to realize this assumption is to share a \textit{pre-determined} generator between the server and the devices, which generates different permutation matrices for different inputs. Then each device only needs to send the selected input of the generator, once at the beginning of the training process. 
    	} the local gradient vector sent from device $k$ as 
    	\begin{align}\label{eq:grad_est2}
        	\hat{\bf g}_k[t] = \sqrt{\frac{\|{\bf g}_k{(t)}\|^2}{N_{\rm w}}}{\bf P}_k^{\sf T} \hat{\bf x}_k[t].
    	\end{align}
    	Our strategy in \eqref{eq:transmitted_signal2} implies that the local gradient elements transmitted at each radio resource are associated with not only different devices but also different weights of the neural network. Therefore, when the local gradient vectors are sparse, it is likely that only a small number of devices simultaneously transmit non-zero local gradient elements at each radio resource, which results in the sparsity of the transmitted signal ${\bf x}[t,n]$.
    	Fig.~\ref{fig:Transmit} illustrates the transmitted signals and the corresponding received signals when employing our random permutation strategy.

	\begin{figure*}[t]
		\centering 
		{\epsfig{file=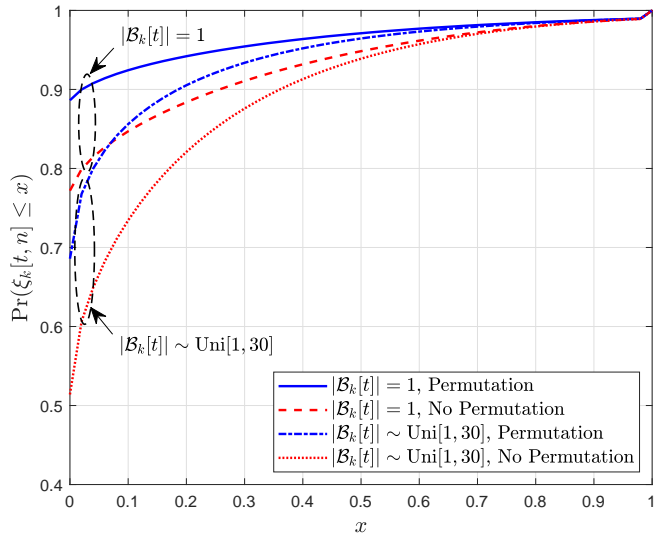, width=10cm}}
		\caption{Cumulative distribution function of the magnitude ratio $\xi_{k}[t,n]$ with and without our random permutation strategy when $M=25$ and $K=100$.} \vspace{-7mm}
		\label{fig:Sparsity}
	\end{figure*}

	    We also demonstrate the sparsity of the transmitted signal through numerical simulations. As a performance metric to evaluate the sparsity, we consider a \textit{magnitude ratio} $\xi_{k}[t,n]$ defined as the ratio of each element's magnitude to the maximum magnitude in the transmitted signal ${\bf x}[t,n]$, i.e., $\xi_{k}[t,n] \triangleq \frac{|x_k[t,n]|}{\max_{k^\prime} |x_{k^{\prime}}[t,n]|}$. Fig.~\ref{fig:Sparsity} shows the cumulative distribution function of $\xi_{k}[t,n]$ with and without our random permutation strategy  when $M=25$ and $K=100$ for the simulation setting described in Sec.~\ref{Sec:Setting}. It should be noticed that the distribution function at $\xi_{k}[t,n]=0$ represents the ratio of the number of zero elements to the  number of total elements in each transmitted signal. It is observed that our random permutation strategy significantly increases the number of zero elements in the transmitted signal for both the stochastic setting ($|\mathcal{B}_k[t]|=1$) and the mini-batch setting ($|\mathcal{B}_k[t]|\sim\textrm{Uni}[1,30]$). Particularly for the stochastic setting, only $10\%$ of the elements of the transmitted signal are non-zero when applying our strategy. Another important observation is that the sparsity level is higher for the stochastic setting than for the mini-batch setting. This result demonstrates that the sparsity of the transmitted signal increases as the batch size decreases, as we already discussed in Sec.~\ref{Sec:Sparsity}.

	\subsection{Reception Strategy: Compressive Sensing}\label{Sec:Proposed}
	    The goal is to design an efficient signal recovery algorithm at the central server to estimate the transmitted signal ${\bf x}[t,n]$ from the received signal ${\bf y}[t,n]$ in \eqref{eq:received_signal}. To achieve this goal, we propose to exploit the sparsity of the transmitted signal by using a compressive sensing approach. The intuition behind this idea is that estimating a \textit{sparse} transmitted signal ${\bf x}[t,n]$ from the received signal ${\bf y}[t,n]= {\bf H}[t,n] {\bf x}[t,n] + {\bf z}[t,n]$ can be interpreted as compressive sensing to estimate an unknown \textit{sparse} signal from its linear measurement with noise \cite{CS:Book}. 
    	Based on this intuition, we propose a compressive sensing algorithm that iteratively finds the LMMSE estimate of the transmitted signal by exploiting its sparsity.
    	This algorithm can be separately applied to the received signal at each radio resource of each uplink phase. Therefore, to simplify the notation, we omit the indexes $t$ and $n$ in the rest of this subsection.

	\vspace{2mm}
	\subsubsection{Definitions}
	We start by defining some terminologies and notations employed in the proposed algorithm. 
	A \textit{true support set} $\tilde{\mathcal{K}} \subset \mathcal{K}$ is a set of device indices that have non-zero gradient values, i.e.,
	\begin{align}\label{eq:true_support}
	    \tilde{\mathcal{K}}\triangleq \{x_k \neq 0~|~k\in\mathcal{K}\}.
	\end{align}
	A \textit{support set} at iteration $i$, denoted by $\mathcal{S}_i \subset \mathcal{K}$, is a set of device indices that have been selected as the members of the true support set until iteration $i$. An estimated transmitted signal at iteration $i$, denoted by $\hat{\bf x}^{(i)} \in \mathbb{R}^{i}$, is the estimate of the transmitted signal when assuming $\tilde{\mathcal{K}} = \mathcal{S}_i$. A \textit{residual vector} at iteration $i$ is defined as ${\bf r}_i ={\bf y}-{\bf H}^{(i)}\hat{\bf x}^{(i)}$ which is a residual part of the received signal after subtracting the effect of the estimated transmitted signal at iteration $i$, where ${\bf H}^{(i)}$ is defined as 
	\begin{align}
	{\bf H}^{(i)}=\big[{\bf h}_{\mathcal{S}_i(1)}, \cdots, {\bf h}_{\mathcal{S}_i(i)}\big].
	\end{align}

	\subsubsection{Support Set Update}
	At iteration $i$, the proposed algorithm selects the index of the device whose normalized channel has the maximum correlation with the residual vector ${\bf r}_{i-1}$ at the previous iteration.
	Our strategy for this step is to use the following selection criterion:
	\begin{align}\label{eq:Criterion}
	    k_i^{\star} = \argmax_{k \in \mathcal{S}_{i-1}^c} \big|\tilde{\bf h}_k^{\sf T} {\bf r}_{i-1}\big|^2,
	\end{align}
	where $\tilde{\bf h}_k = \frac{1}{\|{\bf h}_k\|}{\bf h}_k$. The promising feature of the above criterion is that it correctly finds the device index with the maximum effective SNR, defined as $\rho_k = \frac{1}{\sigma^2}\|{\bf h}_k\|^2|x_k|^2$, when the number of the server's antennas is sufficiently large. This fact can be readily shown by characterizing the correlation between the channel vector of device $k$ and the residual vector, given by  
	\begin{align}\label{eq:Criterion_corr}
    	\frac{1}{M}{\bf h}_k^{\sf T} {\bf r}_{i-1} 
    	&= \frac{1}{M}{\bf h}_k^{\sf T}\big({\bf y}-{\bf H}^{(i-1)} \hat{\bf x}^{(i-1)}\big) \nonumber \\
    	&=\frac{\|{\bf h}_k\|^2}{M}x_k + \sum_{s^\prime \in \mathcal{S}_{i-1}^c\setminus\{k\}} 
    	\frac{{\bf h}_k^{\sf T}{\bf h}_{s^\prime}}{M} x_{s^\prime} + \sum_{s\in \mathcal{S}_{i-1}}\frac{{\bf h}_k^{\sf T}{\bf h}_s}{M} 
    	\big(x_s - \hat{x}_s\big) + \frac{{\bf h}_k^{\sf T}{\bf z}}{M},
	\end{align}
	where $\hat{\bf x}^{(i-1)}=\big[\hat{x}_{\mathcal{S}_{i-1}(1)}, \cdots, \hat{x}_{\mathcal{S}_{i-1}(i-1)}\big]^{\sf T}$.
	By the law of large numbers, the correlation in \eqref{eq:Criterion_corr} approaches $x_k$ as $M$ increases; thereby, the metric in \eqref{eq:Criterion} can be approximated as 
	\begin{align}\label{eq:Criterion_approx}
	\big|\tilde{\bf h}_k^{\sf T} {\bf r}_{i-1}\big|^2 \approx \|{\bf h}_k \|^2 |x_k|^2 \propto \rho_k,
	\end{align}
	for $M\gg 1$. Since we focus on the massive MIMO system with $M\gg 1$, the proposed algorithm at iteration $i$ is expected to find the $i$-th dominant element in the true support set $\tilde{\mathcal{K}}$. Once the best index $k_i^\star$ is selected by the criterion in \eqref{eq:Criterion}, the support set $\mathcal{S}_{i-1}$ is updated by adding the selected index, i.e.,
	\begin{align}\label{eq:Set_update}
	    \mathcal{S}_i\leftarrow \mathcal{S}_{i-1}\cup \{k_i^\star\}.
	\end{align}


	\subsubsection{Transmitted Signal Estimation}
	After updating the support set at iteration $i$, the proposed algorithm estimates the transmitted signal associated with the current support set $\mathcal{S}_i$. 
	Our strategy for this step is to find the LMMSE estimate that has the minimum MSE with respect to the true transmitted signal. For a given support set $\mathcal{S}_i$, the received signal at the server can be rewritten as 
	\begin{align}\label{eq:y_iter}
	    {\bf y} =  \sum_{s\in\mathcal{S}_{i}}  {\bf h}_s x_s  + \sum_{s^\prime \in\mathcal{S}_{i}^{c}} {\bf h}_{s^\prime} x_{s^\prime} + {\bf z}.
	\end{align}
	By the support set update rule in \eqref{eq:Criterion}, the first term of the right hand side (RHS) of \eqref{eq:y_iter} may consist of signals with the $i$ largest magnitudes.
	Motivated by this observation, we approximate the received signal in \eqref{eq:y_iter}  by assuming that the magnitude of $\sum_{s\in\mathcal{S}_{i}}  {\bf h}_s x_s $ is much larger than the magnitude of $\sum_{s^\prime \in\mathcal{S}_{i}^{c}} {\bf h}_{s^\prime} x_{s^\prime}$, expressed as
	\begin{align}
	    {\bf y} \approx \sum_{s\in\mathcal{S}_{i}}  {\bf h}_s x_s + {\bf z} ={\bf H}^{(i)}{\bf x}^{(i)} + {\bf z},
	\end{align}
	where ${\bf x}^{(i)}= [x_{k_1^{\star}}, \cdots, x_{k_i^{\star}}]^{\sf T}$. From the above approximation, the LMMSE estimate for ${\bf x}^{(i)}$ is computed as \cite{Tse:Book}
	\begin{align}\label{eq:x_LMMSE}
    	\hat{\bf x}^{(i)} = {\bf R}_{\bf x}^{(i)}({\bf H}^{(i)})^{\sf T} \left( {\bf H}^{(i)}{\bf R}_{\bf x}^{(i)}({\bf H}^{(i)})^{\sf T} + \sigma^2 {\bf I}_{2M}\right)^{-1}({\bf y} - {\bm \mu}_{\bf y}) + {\bm \mu}_{\bf x}^{(i)},
	\end{align}
	provided that $\mathbb{E}[{\bf y}] = {\bm \mu}_{\bf y}$, $\mathbb{E}[{\bf x}^{(i)}] = {\bm \mu}_{\bf x}^{(i)}$, and $\mathbb{E}[({\bf x}^{(i)}-{\bm \mu}_{\bf x}^{(i)})({\bf x}^{(i)}-{\bm \mu}_{\bf x}^{(i)})^{\sf T}] = {\bf R}_{\bf x}^{(i)}$.

	Unfortunately, the LMMSE estimate in \eqref{eq:x_LMMSE} cannot be computed at the server since it does not have the information of the mean and the covariance of the transmitted signal. To overcome this difficulty, we model these statistics based on relevant observations and approximations. 
	First of all, from the large-scale approximation in \eqref{eq:Criterion_approx}, we can approximate $|x_{k_i^\star}|^2$ as
	\begin{align}\label{eq:alpha_approx}
	    |x_{k_i^\star}|^2 \approx \frac{1}{\|{\bf h}_{k_i^{\star}}\|^2}|\tilde{\bf h}_{k_i^{\star}}^{\sf T} {\bf r}_{i-1}|^2 \triangleq \alpha_{k_i^{\star}}.
	\end{align}		
	Note that the above approximation is tight in the massive MIMO system because $\alpha_{k_i^{\star}}$ approaches $|x_{k_i^{\star}}|^2$ as the number of the server's antennas increases, as shown in \eqref{eq:Criterion_approx}.
	Since we only have the information of $|x_{k_i^\star}|^2 \approx \alpha_{k_i^{\star}}$, a simple, yet reasonable, model for $x_{k_i^\star}$ would be a discrete random variable with a probability mass function:
	\begin{align}\label{eq:x_model}
	    \mathbb{P}(x_{k_i^\star}=x) = \begin{cases}   
	    \frac{1}{2}, & x=\sqrt{\alpha_{k_i^{\star}}}, \\
	    \frac{1}{2}, &  x=-\sqrt{\alpha_{k_i^{\star}}}.
	    \end{cases}
	\end{align}	
	Under this model, the mean and the variance of $x_{k_i^\star}$ is given by $\mathbb{E}[x_{k_i^\star}]=0$ and $\mathbb{E}[|x_{k_i^\star}|^2] = \alpha_{k_i^{\star}}$, and consequently, we have ${\bm \mu}_{\bf x}^{(i)} = {\bf 0}_{i}$, ${\bm \mu}_{\bf y}={\bf H}^{(i)}{\bm \mu}_{\bf x}^{(i)} = {\bf 0}_{i}$, and ${\bf R}_{\bf x}^{(i)}=\mathbb{E}[{\bf x}^{(i)}({\bf x}^{(i)})^{\sf T}]$.
	Now, recall from the discussions in Sec.~\ref{Sec:Permute} that the elements of the transmitted signal are statistically uncorrelated (i.e., $\mathbb{E}[x_k x_j] = 0$ for $k\neq j$ with $k,j\in \mathcal{K}$) because they are associated with different parameters by the use of the random permutation strategy. Utilizing this fact, we may assume that all the non-diagonal elements of ${\bf R}_{\bf x}^{(i)}$ are zero, which yields 
	\begin{align}\label{eq:R_approx}
	    {\bf R}_{\bf x}^{(i)} 
	    &= {\sf diag}\big(\mathbb{E}[|x_{k_1^\star}|^2],\mathbb{E}[|x_{k_2^\star}|^2],\ldots,\mathbb{E}[|x_{k_i^\star}|^2]\big) \\
	    &= {\sf diag}\big(\alpha_{k_1^{\star}},\alpha_{k_2^{\star}},\ldots,\alpha_{k_i^{\star}}\big) \triangleq {\bf D}_{\alpha}^{(i)},
	\end{align}
	where ${\sf diag}(a_1,\cdots,a_N)$ is an $N\times N$ diagonal matrix whose $i$-th diagonal element is $a_i$.

	Based on the statistical model discussed above, the proposed algorithm estimates the transmitted signal at iteration $i$ as
	\begin{align}\label{eq:Estimated_signal0}
	\hat{\bf x}^{(i)} = {\bf D}_{\alpha}^{(i)}({\bf H}^{(i)})^{\sf T} 
	    \left( {\bf H}^{(i)}{\bf D}_{\alpha}^{(i)}({\bf H}^{(i)})^{\sf T} + \sigma^2 {\bf I}_{2M}\right)^{-1}{\bf y}.
	\end{align}
	To further reduce the computational complexity, we rewrite the estimate in \eqref{eq:Estimated_signal0} as
	\begin{align}\label{eq:Estimated_signal}
	\hat{\bf x}^{(i)} = {\bf D}_{\alpha}^{(i)}({\bf H}^{(i)})^{\sf T} {\bm \Omega}_i{\bf y},
	\end{align}
	where 
	\begin{align}\label{eq:Omega_def}
	{\bm \Omega}_i = \left( {\bf H}^{(i)}{\bf D}_{\alpha}^{(i)} ({\bf H}^{(i)})^{\sf T} + \sigma^2 {\bf I}_{2M}\right)^{-1}.
	\end{align}
	Then ${\bm \Omega}_i$ in \eqref{eq:Omega_def} can be computed in a recursive manner:
	\begin{align}\label{eq:Omega_update}
	{\bm \Omega}_i 
	&= \left({\bm \Omega}_{i-1}^{-1} + \alpha_{k_i^\star} {\bf h}_{k_i^\star}{\bf h}_{k_i^\star}^{\sf T}\right)^{-1} ={\bm \Omega}_{i-1} - \frac{\alpha_{k_i^\star} {\bm \Omega}_{i-1}{\bf h}_{k_i^\star}{\bf h}_{k_i^\star}^{\sf T}{\bm \Omega}_{i-1}}{1 + \alpha_{k_i^\star}{\bf h}_{k_i^\star}^{\sf T} {\bm \Omega}_{i-1}{\bf h}_{k_i^\star}},
	\end{align}
	where the second equality holds from the matrix inversion lemma \cite{Woodbury}.
	Therefore, the proposed algorithm reduces the computational complexity required for estimating the transmitted signal, by recursively updating ${\bm \Omega}_i$ at each iteration.

	\subsubsection{Stopping Criterion}
	An ideal stopping criterion for the proposed algorithm is very difficult to derive without assuming perfect information of the true support set and the true transmitted signal at the server. As an alternative approach, we design a stopping criterion that is expected to act optimally under an ideal scenario, in which 1) all elements of the true support set are correctly selected during the first $|\tilde{\mathcal{K}}|$ iterations; and 2) the transmitted signal associated with the true support set follows the statistical model assumed in the proposed algorithm, i.e.,  $\mathbb{E}[x_k] = 0$ and $\mathbb{E}[|x_k|^2] = \alpha_k$ for $k\in\mathcal{K}_{\rm true}$. Although this scenario is ideal, it can also be realized in a massive MIMO system because both conditions hold from \eqref{eq:Criterion_approx} and \eqref{eq:R_approx} when the number of the server's antennas is sufficiently large.

	For the ideal scenario discussed above, we design a stopping criterion by deriving an analytical threshold for the norm on the residual vector. In this scenario, the support set at iteration $i$ belongs to one of the following cases: 
	\begin{itemize}
		\item {\bf Case 1:} The support set at iteration $i$ is a subset of the true support set, but not equal to the true set, i.e., $\mathcal{S}_i \subset \tilde{\mathcal{K}}$ and $\mathcal{S}_i \neq \tilde{\mathcal{K}}$. 
		\item {\bf Case 2:} The support set at iteration $i$ is equal to the true support set, i.e., $\mathcal{S}_i =\tilde{\mathcal{K}}$. 
		\item {\bf Case 3:} The support set at iteration $i$ includes the true support set and has one more element than the true set, i.e., $\mathcal{S}_{i} = \tilde{\mathcal{K}} \cup \{k_i^\star\}$.
	\end{itemize}
	Clearly, the optimal decision for the proposed algorithm is to stop if the current set belongs to {\bf Case 2}. Motivated by this, we derive a condition that determines the case to which the current support set belongs. To achieve this goal, we characterize the expected value of the norm squared of the residual vector in these three cases. The result of this characterization is given in the following proposition:
	\begin{prop}
		If the support set at iteration $i$ belongs to {\bf Case $p$}, the expected value of the norm squared of the current residual vector is given by 
		\begin{align}
		\mathbb{E}[\|{\bf r}_i\|^2] = E_p^{(i)},
		\end{align}
		for $p\in\{1,2,3\}$,
		where
		\begin{align}
		E_1^{(i)} &=\sigma^4\left[{\sf Tr}({\bm \Omega}_i) + 
		\sum_{k \in\tilde{\mathcal{K}}\setminus\mathcal{S}_i} \alpha_k \|{\bm \Omega}_i{\bf h}_k\|^2\right],\\
		E_2^{(i)} &=\sigma^4{\sf Tr}({\bm \Omega}_i), \\ 
		E_3^{(i)} &= \sigma^4\left[{\sf Tr}({\bm \Omega}_i) - 
		\frac{{\sf Tr}({\bm \Omega}_{i-1}) - {\sf Tr}({\bm \Omega}_i)}
		{ 1+ \alpha_{k_i^\star}{\bf h}_{k_i^\star}^{\sf T} {\bm \Omega}_{i-1}{\bf h}_{k_i^\star}}\right],
		\end{align}    
		provided that $\mathbb{E}[x_k] = 0$ and $\mathbb{E}[|x_k|^2] = \alpha_k$ for $k\in\tilde{\mathcal{K}}$. 
	\end{prop}
	\begin{IEEEproof}
		See Appendix A. 
	\end{IEEEproof}
	\vspace{1mm}
	\noindent Proposition~1 shows that $\mathbb{E}[\|{\bf r}_i\|^2]$ decreases as the algorithm proceeds, while $E_1^{(i)} \geq E_2^{(i)} \geq E_3^{(i)}$.  Therefore, the current support set is expected to belong to {\bf Case 3} if the norm squared of the residual vector is closer to $E_3^{(i)}$ than to $E_2^{(i)}$. Utilizing this observation, we set the stopping criterion of the proposed algorithm as $\|{\bf r}_i\|^2 \leq E_{\rm th}^{(i)}$, where 
	\begin{align}\label{eq:Error_th}
	E_{\rm th}^{(i)} 
	&= \frac{1}{2}\big(E_2^{(i)} + E_3^{(i)}\big) = {\sigma^4}\left[{\sf Tr}({\bm \Omega}_i) - 
	\frac{{\sf Tr}({\bm \Omega}_{i-1}) - {\sf Tr}({\bm \Omega}_i)}
	{ 2\big(1+ \alpha_{k_i^\star}{\bf h}_{k_i^\star}^{\sf T} {\bm \Omega}_{i-1}{\bf h}_{k_i^\star}\big)}\right].     
	\end{align}
	The above criterion checks whether the support set at iteration $i$ belongs to {\bf Case 3} or not. This is equivalent to checking whether the support set at the previous iteration $i-1$ belongs to {\bf Case 2} or not. Therefore, after the algorithm stops by satisfying $\|{\bf r}_i\|^2 \leq E_{\rm th}^{(i)}$, the final estimate of the transmitted signal is set as the previous estimate obtained at iteration $i-1$, instead of the current estimate.

	\begin{algorithm}
		\caption{Federated learning over a massive MIMO system with the presented compressive sensing approach.}\label{alg:RL}
		{\small{\begin{algorithmic}[1]
					\STATE Initialize the weight vector ${\bf w}_1$. 
					\FOR {$t=1$ to $T$} 
					\STATE \textit{At the server:}
					\STATE ~~~Transmit ${\bf w}_{t}$ to $M$ wireless devices.
					\STATE \textit{At device $k\in\mathcal{K}$:}
					\STATE ~~~Compute ${\bf g}_k[t]$ from \eqref{eq:local_grad}.
					\STATE ~~~Compute ${\bf x}_k[t]$ from \eqref{eq:transmitted_signal2}.
					\STATE ~~~Transmit ${\bf x}_k[t]$ to the server.
					\STATE \textit{At the server:}
					\FOR {$n=1$ to $N_{\rm w}$}	
					\STATE Set ${\bf h}_k = {\bf h}_k[t,n]$ and $\tilde{\bf h}_k = \frac{{\bf h}_k}{\|{\bf h}_k\|}$ for $k\in\mathcal{K}$.  
					\STATE Set $\mathcal{S}_0 = \emptyset$, ${\bf r}_0 = {\bf y}[t,n]$, and ${\bm \Omega}_{0}=\frac{1}{\sigma^2}{\bf I}_{2M}$.
					\FOR {$i=1$ to $K$}
					\STATE Find $k_i^\star = \argmax_{k \in \mathcal{S}_{i-1}^c} \big|\tilde{\bf h}_k^{\sf T} {\bf r}_{i-1}\big|^2$. 
					\STATE Set $\mathcal{S}_i = \mathcal{S}_{i-1}\cup \{k_i^\star\}$ and $\alpha_{k_i^\star} = \frac{1}{\|{\bf h}_{k_i^\star}\|^2}\big|\tilde{\bf h}_{k_i^\star}^{\sf T} {\bf r}_{i-1}\big|^2$. 
					\STATE Compute ${\bm \Omega}_i$ from \eqref{eq:Omega_update}.  
					\STATE Set ${\bf r}_i = \sigma^2{\bm \Omega}_i{\bf y}[t,n]$.  
					\STATE Compute $E_{\rm th}^{(i)}$ from \eqref{eq:Error_th}. 
					\IF{$\|{\bf r}_i\|^2 < E_{\rm th}^{(i)}$}   
					\STATE Update $I^\star = i-1$.
					\STATE Break the loop.
					\ENDIF
					\ENDFOR
					\STATE Compute $\hat{\bf x}^{(I^\star)} = {\bf D}_{\alpha}^{(I^\star)}({\bf H}^{(I^\star)})^{\sf T} {\bm \Omega}_{I^\star}{\bf y}[t,n]$. 
					\STATE Set $\hat{x}_{k_i^\star}[t,n] = \hat{x}_i^{(I^\star)}$ for $i\in\{1,\ldots,I^\star\}$.
					\STATE Set $\hat{x}_k[t,n] = 0$ for $m\notin\mathcal{S}_i$.
					\ENDFOR
					\STATE Compute $\hat{\bf g}_k[t]$ from \eqref{eq:grad_est2} for $k\in\mathcal{K}$.
					\STATE Compute $\bar{\bf g}[t]$ from \eqref{eq:global_grad}. 
					\STATE Update ${\bf w}_{t+1}$ based on $\bar{\bf g}[t]$.
					\ENDFOR
		\end{algorithmic}}}
	\end{algorithm}
	
	\subsubsection{Summary}
	In Algorithm~1, we summarize the overall process of federated learning over the massive MIMO system when employing the presented compressive sensing approach. 
	In this algorithm, Steps $3{\sim}4$ and Steps $5{\sim}30$ are associated with the downlink and uplink phases, respectively.
	Note that Steps $11{\sim}26$ may change according to the reception strategy adopted by the central server. In Step $17$, the residual vector ${\bf r}_i$ is determined using the following equality:
	\begin{align}\label{eq:Residual_simple}
	{\bf r}_i &= {\bf y}-{\bf H}^{(i)}\hat{\bf x}^{(i)} = \Big({\bf I}_{2M}-{\bf H}^{(i)}{\bf R}_{\bf x}^{(i)}({\bf H}^{(i)})^{\sf T} {\bm \Omega}_i \Big){\bf y}= \sigma^2{\bm \Omega}_i{\bf y},
	\end{align}
	where the second equality is obtained from \eqref{eq:Estimated_signal}. By using the expression in \eqref{eq:Residual_simple}, computing the estimated transmitted signal $\hat{\bf x}^{(i)}$ is not required at each iteration. Instead, $\hat{\bf x}^{(i)}$ is computed only once after the proposed algorithm ends (see Step 23). Using this strategy, the overall computational complexity of the proposed algorithm is further reduced.
        As can be seen in Algorithm~1, the proposed algorithm is a greedy algorithm whose optimality is not guaranteed in general. Nevertheless, the numerical results in Sec.~V illustrate that the performance gap between federated learning with the proposed algorithm and centralized learning with perfect reconstruction is marginal under certain scenarios.

	\vspace{2mm}
    	{\bf Remark} (\textit{Comparison to Conventional Orthogonal Matching Pursuit Algorithms}): 
    	The proposed algorithm can be regarded as a variant of the orthogonal matching pursuit (OMP) algorithm in compressive sensing. The common feature of the proposed algorithm and the conventional OMP algorithms in \cite{Tropp:04,Cai:11,Sundin:13,Sparrer:16} is that a support set is iteratively updated by adding an index with the maximum correlation to the residual vector. Despite this fact, the proposed algorithm still differs from the algorithms in \cite{Tropp:04,Cai:11,Sundin:13,Sparrer:16}. The major differences between the proposed and the conventional OMP algorithms are summarized as follows:  
	\begin{itemize}
		\item The proposed algorithm utilizes the LMMSE estimate that generalizes the least-squares (LS) estimate considered in the conventional OMP algorithms in \cite{Tropp:04,Cai:11}. It is well-known that the LMMSE estimate is superior to the LS estimate because the LMMSE estimate is optimal in terms of the MSE. Although the LMMSE estimate is also considered in \cite{Sundin:13,Sparrer:16}, the LMMSE estimate in the proposed algorithm utilizes a different statistical model for the transmitted signal, established based on both a large-scale approximation and a statistical feature obtained from the random permutation strategy.  
		\item The stopping criterion of the proposed algorithm differs from those of the conventional OMP algorithms. Particularly, the criterion of the proposed algorithm is uniquely designed by deriving an analytical threshold for the residual error at each iteration, based on the statistical model established for the transmitted signal in federated learning.
	\end{itemize}

	\section{Comparison to A Linear Beamforming Approach}\label{Sec:BF}
    	Another possible solution for reconstructing the local gradient vectors at the central server is to apply conventional linear beamforming methods developed to solve a MIMO data detection problem \cite{Tse:Book}. 
    	Motivated by this, we introduce two linear beamforming methods as performance benchmarks for the proposed algorithm. 
    	We also discuss the limitation of each method for the use in federated learning over a massive MIMO system. 
    	We then compare the computational complexity of the proposed algorithm with those of the linear beamforming methods. 
	
	
	
	\subsection{Limitation of  Maximal Ratio Combining (MRC)}\label{Sec:MRC} 
	The simplest yet effective linear beamforming method is the MRC method which aims to maximize the power of the desired signal by aligning the direction of the receive beamforming into the channel direction. 
	The estimate of the transmitted signal is given by
	\begin{align}\label{eq:MRC}
	    \hat{\bf x}_{\rm MRC}[t,n] = {\sf diag}\left(\frac{1}{\|{\bf h}_{1}[t,n]\|^{2}},~\ldots~,\frac{1}{\|{\bf h}_{K}[t,n]\|^{2}}\right){\bf H}^{\sf T}[t,n] {\bf y}[t,n].
	\end{align}
	Consequently, the estimate of the local gradient vector is given by $\hat{\bf g}_k^{\rm MRC}[t] = \sqrt{\frac{\|{\bf g}_k{(t)}\|^2}{N_{\rm w}}} \hat{\bf x}_k^{\rm MRC}[t]$, where $\hat{\bf x}_k^{\rm MRC}[t] = \big[ \hat{x}_k^{\rm MRC}[t,1],\cdots,\hat{x}_k^{\rm MRC}[t,N_{\rm w}] \big]^{\sf T}$ and $\hat{x}_k^{\rm MRC}[t,n]$ is the $k$-th element of  $\hat{\bf x}_{\rm MRC}[t,n]$.

    	To highlight a limitation of the MRC method, we take a closer look at the $k$-th element of the estimated signal $\hat{\bf x}_{\rm MRC}[t,n]$ given by 
    	\begin{align}\label{eq:x_hat_MRC_k}
    	\hat{x}_{{\rm MRC},k}[t,n] = x_k[t,n] +  \sum_{j\neq k} \frac{{\bf h}_{k}^{\sf T}[t,n]{\bf h}_j[t,n]}{\|{\bf h}_k[t,n]\|^2} x_j[t,n] + \frac{{\bf h}_{k}^{\sf T}[t,n]{\bf z}[t,n]}{\|{\bf h}_k[t,n]\|^2}.
    	\end{align}
    	As can be seen in \eqref{eq:x_hat_MRC_k}, the estimate in the MRC method consists of not only the desired signal $x_k[t,n]$, but also an inter-user interference (IUI) and an effective noise corresponding to the second and the third terms in the RHS of \eqref{eq:x_hat_MRC_k}, respectively. Although both IUI and noise terms vanish as the number of the server's antennas goes to infinity by the central limit theorem \cite{Random:Book}, this does not hold for a general number of the server's antennas. Therefore, in most practical scenarios, the MRC method is suboptimal in terms of the estimation performance as will be shown in Sec.~\ref{Sec:Simul}.

	\subsection{Limitation of  Linear Minimum Mean Square Error (LMMSE)}\label{Sec:LMMSE} 
	The LMMSE method is an optimal linear beamforming method that minimizes the MSE between the true transmitted signal and its estimate. 
	The estimate of the transmitted signal in the LMMSE method is given by \cite{Tse:Book}
	\begin{align}\label{eq:LMMSE}
    	\hat{\bf x}_{\rm LMMSE}[t,n] = {\bf F}_{\rm LMMSE}[t,n]({\bf y}[t,n]-{\bm \mu}_{\bf y}[t,n]) + {\bm \mu}_{\bf x}[t,n],
	\end{align}
	where 
	\begin{align}\label{eq:LMMSE_BF}
	    {\bf F}_{\rm LMMSE}[t,n] = {\bf R}_{\bf x}[t,n]{\bf H}^{\sf T}[t,n]\left({\bf H}[t,n]{\bf R}_{\bf x}[t,n]{\bf H}^{\sf T}[t,n] + \sigma^2{\bf I}_{2M}\right)^{-1},
	\end{align}
	where ${\bm \mu}_{\bf x}[t,n] = \mathbb{E}[{\bf x}[t,n]]$, ${\bf R}_{\bf x}[t,n] = \mathbb{E}[{\bf x}[t,n]{\bf x}^{\sf T}[t,n]]$, and ${\bm \mu}_{\bf y}[t,n] = \mathbb{E}[{\bf y}[t,n]]$. Consequently, the estimate of the local gradient vector is given by 
	$\hat{\bf g}_k^{\rm LMMSE}[t] = \sqrt{\frac{\|{\bf g}_k{(t)}\|^2}{N_{\rm w}}} \hat{\bf x}_k^{\rm LMMSE}[t]$, where $\hat{\bf x}_k^{\rm LMMSE}[t] = \big[ \hat{x}_k^{\rm LMMSE}[t,1],\cdots,\hat{x}_k^{\rm LMMSE}[t,N_{\rm w}] \big]^{\sf T}$ and $\hat{x}_k^{\rm LMMSE}[t,n]$ is the $k$-th element of  $\hat{\bf x}_{\rm LMMSE}[t,n]$.

	    A major limitation of the LMMSE method is that it is applicable only when both the mean $\mathbb{E}[{\bf x}[t,n]]$ and the covariance ${\bf R}_{\bf x}[t,n]$ of the transmitted signal are known at the server. Characterizing the statistical behavior of the gradient vector is very challenging in the most learning tasks, which is  due to the randomness and the heterogeneity of real-world data. For this reason, the statistics of the transmitted signal are generally unknown at the server, and consequently, the LMMSE method suffers from performance degradation caused by imperfect statistical information. Another limitation of the LMMSE method is that it requires the computation of a $2M\times 2M$ matrix inversion to determine the beamforming matrix in \eqref{eq:LMMSE_BF}. Therefore, the computational complexity required in this method may not be affordable in the massive MIMO system with a large number of the server's antennas, as will be discussed in detail in the following subsection.

	\begin{table*}
		\renewcommand{\arraystretch}{1.7}
		\caption{The number of real multiplications required for various estimation methods.}
		\label{table:complexity}
		\centering
		\begin{tabular}{c||c|c}
			\hline
			\bfseries Method &  General case & Large-scale case ($M\gg 1$ and $K\gg 1$) \\
			\hline\hline
			Proposed 
			& ~~\makecell{ $(I^\star)^2 \big(M+\frac{1}{2}\big) + I^\star \big(12M^2 + 2KM + 11M + K +\frac{17}{2}\big) $ \\  $ +12M^2 + 4KM + 10M + 2K + 7$ }~~
			& ~~\makecell{ $(I^\star)^2 M + I^\star (12M^2 + 2KM) $ \\  $ + 12 M^2 + 4KM$}~~
			\\ 			\hline	
			MRC 
			& $4KM + K$ & $4KM$ 
			\\			\hline
			LMMSE 
			& $8KM^2 -4M^2+ 6KM + 2M$  
			& $8KM^2 $ 
			\\			\hline
		\end{tabular}
	\end{table*}	
	
	\subsection{Comparison of Computational Complexity}\label{Sec:Complex} 
	We analyze and compare the computational complexity of the proposed algorithm in Sec.~\ref{Sec:Proposed}, the MRC method in Sec.~\ref{Sec:MRC}, and the LMMSE method in Sec.~\ref{Sec:LMMSE}.
	    To this end, we count the number of real multiplications required to compute Steps 13--24 in Algorithm~\ref{alg:RL} for the proposed algorithm, \eqref{eq:MRC} for the MRC method, and \eqref{eq:LMMSE} for the LMMSE method.
    	Particularly for the LMMSE method, we present the minimum complexity by considering a recursive computation of matrix inversion, as done in \eqref{eq:Omega_update}, under the assumptions of ${\bm \mu}_{\bf x}[t,n]={\bf 0}_K$ and ${\bf R}_{\bf x}[t,n]={\bf I}_K$.
    	The complexity results for three methods are summarized in Table~\ref{table:complexity}, where $I^\star$ is the size of the support set determined by the proposed algorithm (see Algorithm~1). 
	In Table~\ref{table:complexity}, we also present the complexity for a large-scale scenario (i.e., $M\gg 1$ and $K\gg 1$) which is the region of interest in massive MIMO systems.

	Table~\ref{table:complexity} shows that the proposed algorithm has a significantly lower complexity compared to the LMMSE method, when the transmitted signal is very sparse. More precisely, the ratio of the complexity of the proposed algorithm to that of the LMMSE method is obtained as 
	\begin{align}\label{eq:Complex_ratio}
	\frac{C_{\rm Pro}^{\rm large}}{C_{\rm LMMSE}^{\rm large}} 
	&=  \frac{(I^\star)^2}{8KM} + \frac{3I^\star }{2K} +\frac{I^\star}{4M},
	\end{align}
	for $M\gg 1$ and $K\gg 1$.
	If the proposed algorithm properly stops with $I^\star=|\tilde{\mathcal{K}}|$, the complexity ratio in \eqref{eq:Complex_ratio} becomes 
	\begin{align}\label{eq:Complex_ratio2}
	\frac{C_{\rm Pro}^{\rm large}}{C_{\rm LMMSE}^{\rm large}}  = \frac{|\tilde{\mathcal{K}}|}{K} \left(\frac{|\tilde{\mathcal{K}}|}{8M} + \frac{3}{2} \right) +\frac{|\tilde{\mathcal{K}}|}{4M},
	\end{align}
	and consequently,
	\begin{align}\label{eq:Complex_ratio3}
	\frac{C_{\rm Pro}^{\rm large}}{C_{\rm LMMSE}^{\rm large}} \rightarrow \frac{|\tilde{\mathcal{K}}|}{4M}~~\text{as}~~\frac{|\tilde{\mathcal{K}}|}{K}\rightarrow 0,
	\end{align}
	for a fixed $M$.
	The above result implies that if the size of the true support set is much smaller than the number of devices, the complexity reduction achieved by the proposed algorithm over the LMMSE method increases with the number of the server's antennas. Therefore, in federated learning over the massive MIMO system, the proposed algorithm is significantly more beneficial than the LMMSE method in terms of the computational complexity. Note that although the MRC method achieves the lowest complexity among three methods, it suffers from a performance degradation as will be shown in Sec.~\ref{Sec:Simul}.

	\begin{figure}[t]
		\centering
		{\epsfig{file=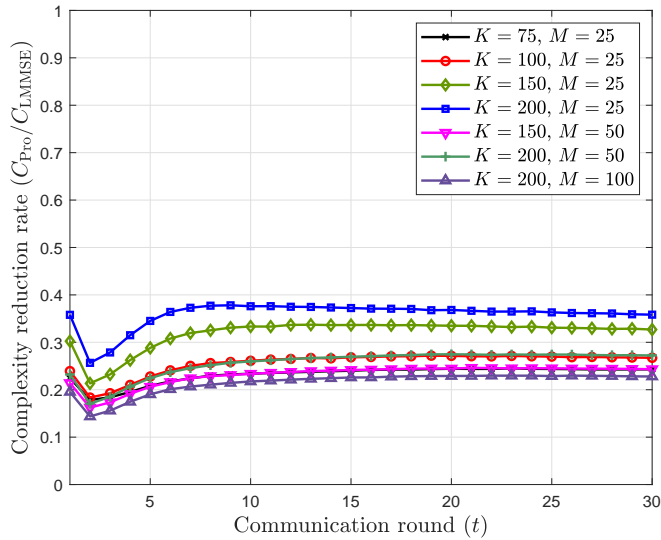, width=10cm}}\vspace{-3mm}
		\caption{The complexity ratio of the proposed algorithm to the LMMSE method for various $K$ and $M$.} \vspace{-3mm}
		\label{fig:Complexity}
	\end{figure}

	    We also compare the computational complexity of the proposed algorithm and the LMMSE method using simulation. In this comparison, we numerically obtain the average size of the support set for the proposed algorithm, namely $I_{\rm avg}^\star$. We then compute the ratio of the number of real multiplications of the proposed algorithm to that of the LMMSE method by using the results in Table~\ref{table:complexity} with $I^\star=I_{\rm avg}^\star$. Fig.~\ref{fig:Complexity} illustrates the complexity ratio of the proposed algorithm to the LMMSE method for different $K$ and $M$ under simulation setting described in Sec.~\ref{Sec:Setting}, where we consider the stochastic setting ($|\mathcal{B}_k[t]|=1$). Fig.~\ref{fig:Complexity} shows that the complexity reduction achieved by the proposed algorithm over the LMMSE method is more than $60\%$ for all the cases. Furthermore, this complexity reduction is shown to be even higher for the case of $K=200$ and $M=100$, corresponding to a large-scale scenario. Therefore, these results demonstrate that the proposed algorithm has a significantly lower complexity compared to LMMSE for the scenarios under consideration. 
	\section{Simulation Results}\label{Sec:Simul}
	In this section, using simulations, we evaluate the classification accuracy of federated learning over a massive MIMO system with various local gradient reconstruction approaches.	
	The wireless channel of the communication system is modeled by $10$-tap CIR that follows uniform power delay profile, in which each CIR tap is distributed as $\mathcal{CN}(0,0.1)$. The number of subcarriers for OFDM signaling is set as $N_{\rm sub}=1024$, and the noise power is set as $\sigma_{\rm c}^2=1$ (i.e., $\sigma^2 =0.5$). 
    	For the implementation of the LMMSE method in Sec.~\ref{Sec:LMMSE}, we assume\footnote{The assumption of ${\bm \mu}_{\bf x}[t,n] = {\bf 0}_{K}$ is employed because we do not have any prior information about the true mean. Also, the assumption of $\mathbb{E}[{\bf x}[t,n]{\bf x}^\top[t,n]] = {\bf I}_K$ is employed because the definition of ${\bf x}_k[t]$ in \eqref{eq:transmitted_signal} implies that $\mathbb{E}[|{x}_k[t,n]|^2] = 1$ if the power of the local gradient vector ${\bf g}_k$ is uniformly distributed across all its elements.} 
        that ${\bm \mu}_{\bf x}[t,n] = {\bf 0}_{K}$ and ${\bf R}_{\bf x}[t,n]={\bf I}_K$.

	\subsection{Simulation Setting}\label{Sec:Setting}
    	In this simulation, we consider an image classification task in which a neural network is used to classify a $28\times 28$ grayscale image of a handwritten digit (from $0$ to $9$) in the MNIST dataset that consists of $60000$ training and $10000$ test data samples \cite{MNIST}. 
        We assume that a central server utilizes a neural network\footnote{Recall that the amount of radio resources required to convey a parameter vector is proportional to the number of the weights of the neural network. Considering this fact, we adopt a relatively simple neural-network structure as it is more suitable for the use in practical communication systems whose radio resources are scarce. Nevertheless, the proposed algorithm can be applied to a more complex neural network (e.g., a convolutional neural network) to further improve the classification accuracy.}
    	that consists of $784$ input nodes, a single hidden layer with $20$ hidden nodes, and $10$ output nodes. The activation functions of the hidden layer and the output layer are set as the ReLU and the softmax functions, respectively. 
    	The weights of the neural network at communication round $t$ are mapped into a parameter vector ${\bf w}_{t}$ with length $N_{\rm w}=15910$. 
    	To train ${\bf w}_{t}$, we adopt the ADAM optimizer in \cite{ADAM} whose update rule is given by
    	\begin{align}
    	    {\bf m}_{t+1} &\leftarrow \beta_1 {\bf m}_t + (1-\beta_1) \bar{\bf g}[t], \label{eq:w_update_ADAM1} \\
    	    {\bf v}_{t+1} &\leftarrow \beta_2 {\bf v}_t + (1-\beta_2) (\bar{\bf g}[t])^2,  \label{eq:w_update_ADAM2} \\
    	    {\bf w}_{t+1} &\leftarrow {\bf w}_{t} - \alpha  {\hat{\bf m}_{t+1}}/({\sqrt{\hat{\bf v}_{t+1}} + \epsilon}),
    	\end{align}
    	where all the operations are element-wise, and we have ${\bf m}_1={\bf 0}_{N_{\rm w}}$, ${\bf v}_1={\bf 0}_{N_{\rm w}}$, $\hat{\bf m}_{t} = {\bf m}_{t}/(1-\beta_1^{t})$, $\hat{\bf v}_{t} = {\bf v}_{t}/(1-\beta_2^{t})$, $\alpha=0.01$, $\beta_1=0.9$, $\beta_2=0.999$, and $\epsilon=10^{-8}$. The global gradient vector $\bar{\bf g}[t]$ required in \eqref{eq:w_update_ADAM1} and \eqref{eq:w_update_ADAM2}  is the function of the local gradient vectors computed-and-sent by the wireless devices, as can be seen in \eqref{eq:global_grad}. For this, we assume that the central server adopts one of the local gradient reconstruction approaches introduced in our work.
    	The local gradient vector at each wireless device is computed according to \eqref{eq:local_grad} using the cross-entropy loss function. 
    	For the local data set $\mathcal{B}_k$ of device $k$, we select the set of $1000$ training data samples at random among the training samples labeled with digit $d_k= \big\lfloor \frac{k-1}{K/10}\big\rfloor$ in the MNIST dataset. This corresponds to a \textit{non-IID} setting because each device has the information of only one digit. Then device $k$ randomly selects $|\mathcal{B}_k[t]|$ samples from $\mathcal{B}_k$ to compute the local gradient vector at communication round $t$. To determine the batch size at each device, we consider two settings: 1) a \textit{stochastic} setting with $|\mathcal{B}_k[t]|=1$, and 2) a \textit{mini-batch} setting with $|\mathcal{B}_k[t]|\sim {\rm Uni}[1,30]$ where $|\mathcal{B}_k[t]|$ is randomly drawn from a uniform distribution over $[1,30]$, for all $k\in\mathcal{K}$ and $t\in\mathcal{T}$.

	\subsection{Classification Accuracy Results}\label{Sec:SimResults}
	\begin{figure}
		\centering 
		\subfigure[$K=75$ and $M=25$]
		{\epsfig{file=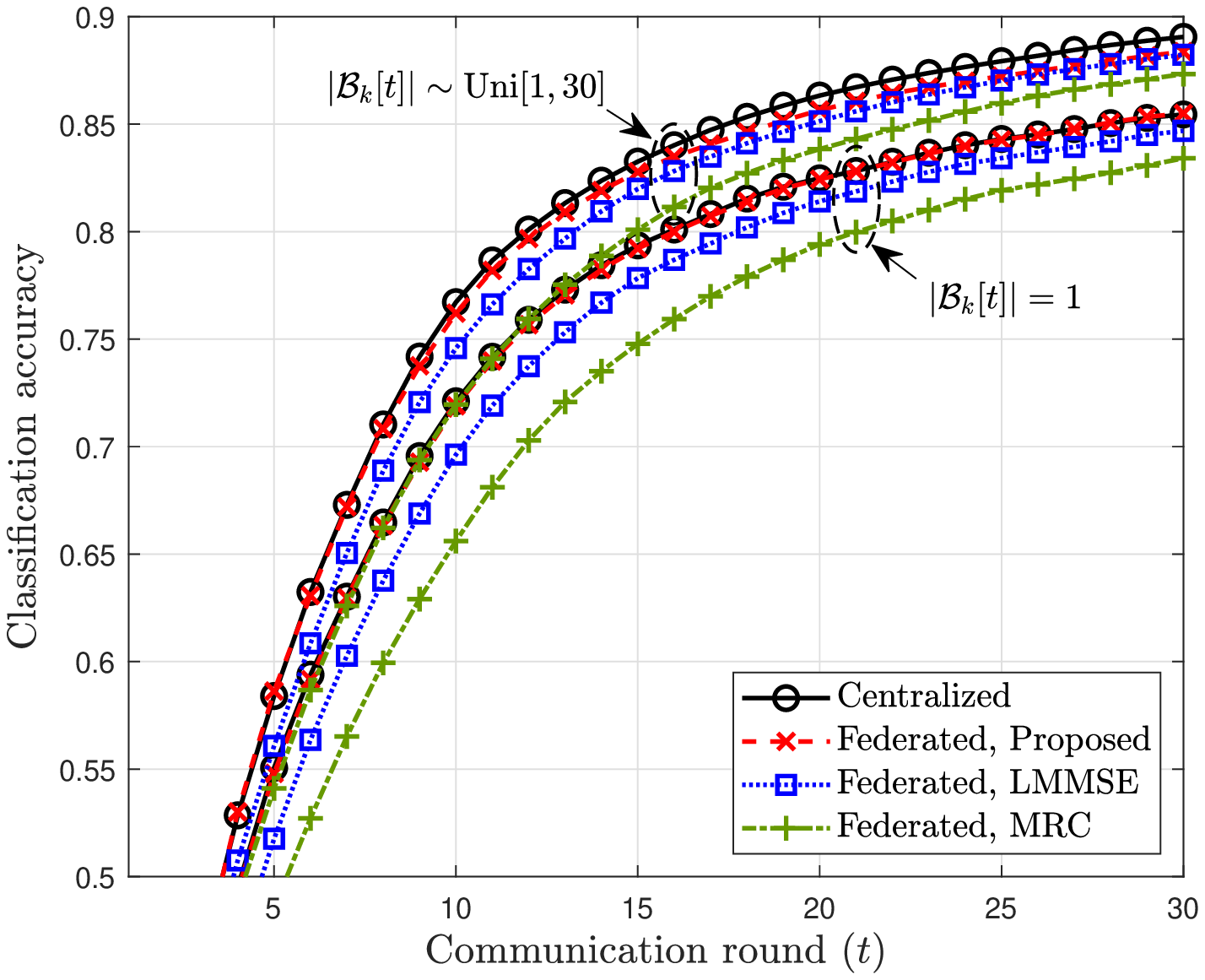, width=8.1cm}}
		\subfigure[$K=150$ and $M=50$]
		{\epsfig{file=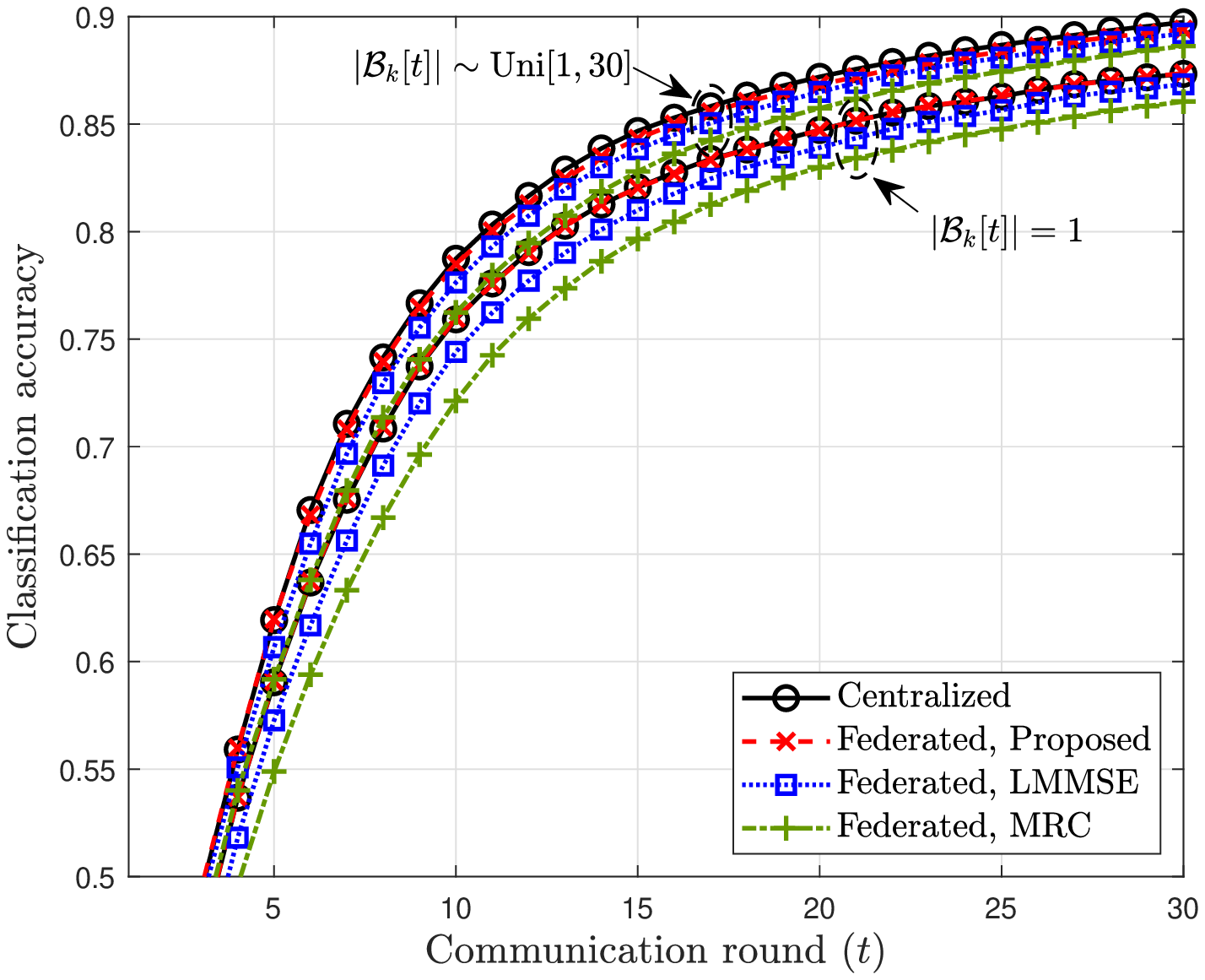, width=8.1cm}}
		\caption{Classification accuracies of federated learning with various local gradient reconstruction approaches for different $K$ and $M$.} \vspace{-3mm}
		\label{fig:Acc}
	\end{figure}
	
    	Fig.~\ref{fig:Acc} compares the classification accuracies of federated learning with various local gradient reconstruction approaches for different $K$ and $M$. As a performance benchmark, we also plot the accuracy achieved by centralized learning with perfect reconstruction of the local gradient vectors at the central server in which $\hat{\bf g}_k[t]={\bf g}_k[t]$ for all $k\in\mathcal{K}$ and $t\in\mathcal{T}$. Both Figs. \ref{fig:Acc}(a) and \ref{fig:Acc}(b) show that the proposed compressive sensing approach outperforms all the linear beamforming methods in terms of the classification accuracy and the convergence rate regardless of the size of the mini-batch data samples used for computing the local gradient vector at each device. In other words, the proposed approach requires a less number of communication rounds than the linear beamforming methods, to achieve the same level of the classification accuracy.
        It is also shown that the performance gap between federated learning with the proposed approach and the centralized learning is marginal for the stochastic setting ($|\mathcal{B}_k[t]|=1$). This result demonstrates that our compressive sensing approach effectively compensates for the performance-degrading factors in wireless communications including IUI, channel fading, and noise effects when the size of the mini-batch samples employed at each device is small.
    	When referring to Figs.~\ref{fig:Complexity} and \ref{fig:Acc} together, it can also be observed that the proposed approach is superior to the LMMSE method in terms of both the classification accuracy and the computational complexity.

	\begin{figure*}[t]
    	\begin{minipage}[b]{8cm}
    		\centering
    		{\epsfig{file=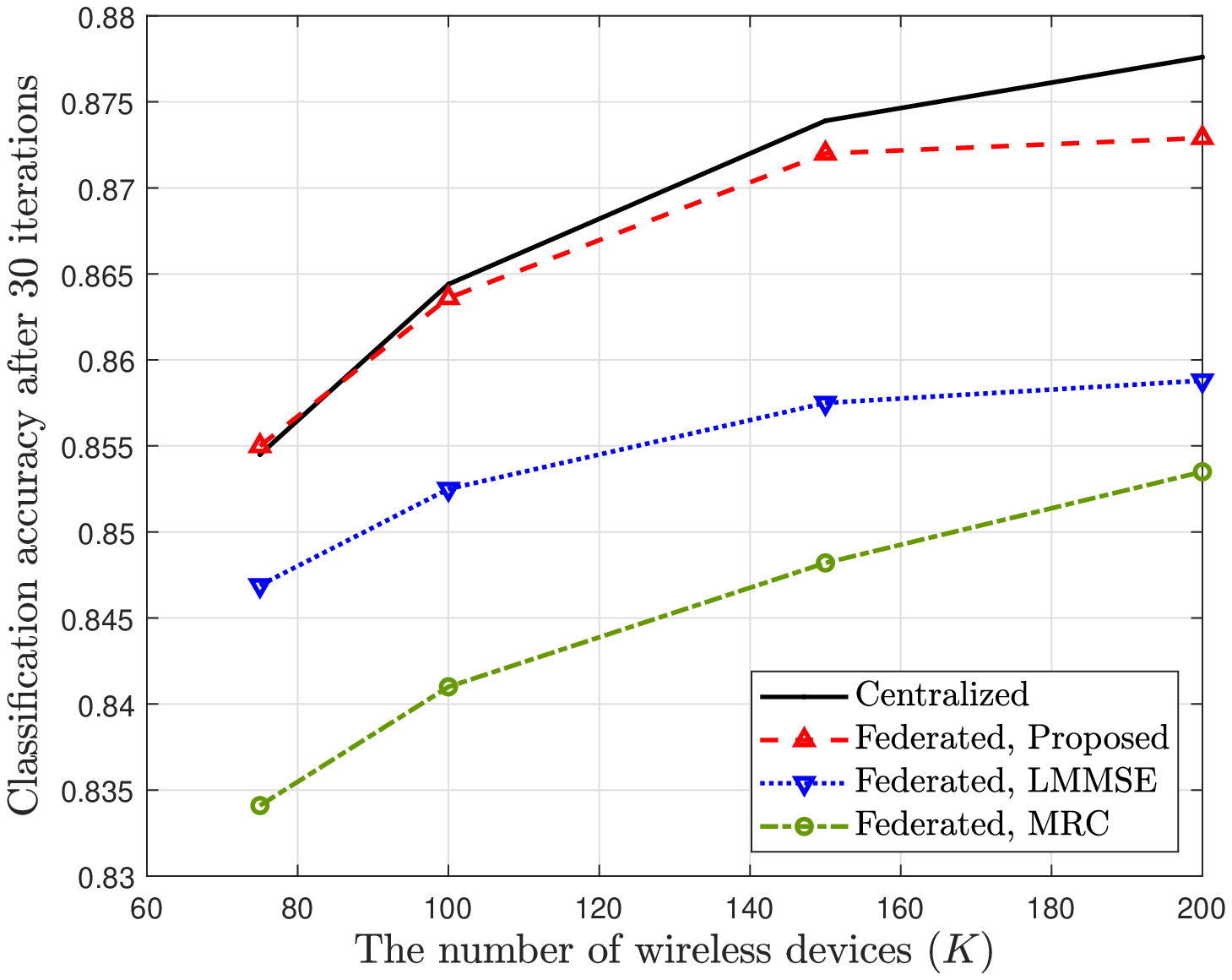, width=8.1cm}}\vspace{-3mm}
    		\caption{The impact of the number of wireless devices $K$ on the classification accuracies after $T=30$ iterations for  federated learning with various local gradient reconstruction approaches when $M=25$ and $|\mathcal{B}_k[t]|=1$.} \vspace{-3mm}
    		\label{fig:UE}
    	\end{minipage}\hfill
    	\begin{minipage}[b]{8cm}
    		\centering
        	{\epsfig{file=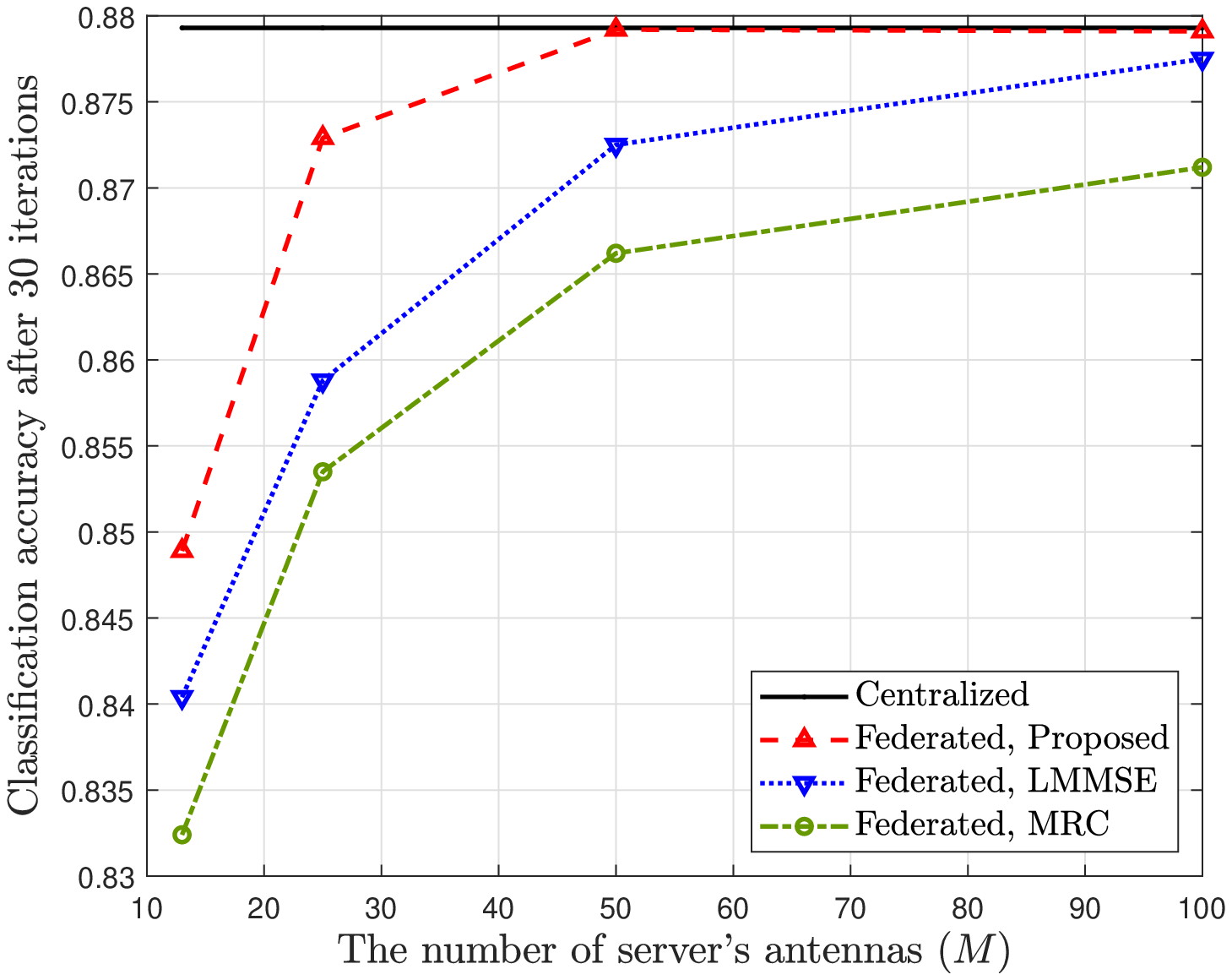, width=8.1cm}}\vspace{-3mm}
        	\caption{The impact of the number of the server's antennas $M$ on the classification accuracies after $T=30$ iterations for  federated learning with various local gradient reconstruction approaches when $K=200$ and $|\mathcal{B}_k[t]|=1$.} \vspace{-3mm}
        	\label{fig:RX}
    	\end{minipage}
    	\vspace{-3mm}
    \end{figure*}
	
	
    	Fig.~\ref{fig:UE} evaluates the impact of the number of wireless devices, $K$, on the classification accuracies after $T=30$ iterations for  federated learning with various local gradient reconstruction approaches when $M=25$ and $|\mathcal{B}_k[t]|=1$. Fig.~\ref{fig:UE} shows that the classification accuracy of all the considered approaches improves with the number of the wireless devices. The intuition behind this result is that increasing the number of the devices leads to an increase in the number of training samples utilized to train the neural network at each communication round. It is also shown that the performance gap between centralized learning and federated learning increases with $K$. The major reason is that the number of unknown values that need to be estimated at the server increases with $K$ while the number of the available observations (i.e., the number of server's antennas) is fixed.  
    	Nevertheless, the proposed compressive sensing approach outperforms all the linear beamforming methods regardless of the number of the devices. This result demonstrates that the performance efficiency of the proposed approach is not degraded by the number of wireless devices participating in the training.
	


    	Fig.~\ref{fig:RX} evaluates the impact of the number of the server's antennas, $M$, on the classification accuracies after $T=30$ iterations for federated learning with various local gradient reconstruction approaches when $K=200$ and $|\mathcal{B}_k[t]|=1$. Fig.~\ref{fig:RX} shows that the classification accuracies of all the considered approaches improves with the number of server's antennas. This performance improvement is the consequence of exploiting the receive diversity that provides the robustness to the performance-degrading factors in wireless communications. Another important observation is that the proposed compressive sensing approach requires a significantly less number of server's antennas than the linear beamforming methods, to achieve the same level of classification accuracy. For example, to achieve $87\%$ classification accuracy, for the proposed approach, the central server requires only 23 antennas, while for LMMSE and MRC, the server should be equipped with 45 and 90 antennas, respectively. Therefore, it is demonstrated that the proposed approach also contributes to reduce the hardware requirement at the server to achieve the desired level of accuracy in federated learning.  
	

	\section{Conclusion}
    	In this paper, we have presented a compressive sensing approach for federated learning over a massive MIMO system, which allows a central server to efficiently reconstruct local gradient vectors sent from wireless devices.
    	In particular, motivated by the sparsity of the local gradient vectors, we have established a proper transmission strategy to construct a sparse transmitted signal that aggregates the signals sent from all the wireless devices at each radio resource.  
    	Based on this transmission strategy, we have proposed a compressive sensing algorithm enabling the central server to iteratively find the LMMSE estimate of the transmitted signal.
    	We have also analyzed the computational complexity of the proposed algorithm and demonstrated that when the transmitted signal is sparse, the proposed algorithm requires a significantly lower complexity compared to the LMMSE method.
    	Using simulations, we have demonstrated that the presented approach outperforms the linear beamforming approaches in terms of the accuracy, while reducing the performance gap between federated learning and centralized learning with perfect reconstruction.
	

	An important direction for future research is to extend our compressive sensing approach by developing a proper device scheduling algorithm. In this extension, a joint optimization of device scheduling and local gradient reconstruction may further improve the performance of federated learning, particularly when the number of wireless devices is much larger than the number of server's antennas. 
	    Another important research direction is to develop a proper downlink transmission strategy for broadcasting the parameter vector to the wireless devices and investigate its impact on the performance of federated learning over a massive MIMO system.
	    It would also be important to provide convergence analysis of federated learning with the proposed approach, in order to shed light on the potentials and limitations of federated learning over practical wireless communication systems. To this end, it would be possible to apply or extend the techniques introduced in \cite{Amiri:Analysis}.

	\appendices
	\section{Proof of Proposition~1}\label{Apdx:Thm1}
	Suppose that $\mathcal{S}_{i-1} \subset \tilde{\mathcal{K}}$, $\mathbb{E}[x_k] = 0$, and $\mathbb{E}[|x_k|^2] = \alpha_k$ for $k\in\tilde{\mathcal{K}}$. 
	Then the norm squared of the residual vector at iteration $i$ is expressed as
	\begin{align}\label{eq:Error_r}
	\mathbb{E}[\|{\bf r}_i\|^2] 
	&= {\sf Tr}\big(\mathbb{E}[{\bf r}_i{\bf r}_i^{\sf T}]\big) = \sigma^4{\sf Tr}\big({\bm \Omega}_i\mathbb{E}[{\bf y}_i{\bf y}_i^{\sf T}]{\bm \Omega}_i\big),
	\end{align}
	where the second equality is obtained from \eqref{eq:Residual_simple}. Thanks to the use of the permutation matrix in \eqref{eq:transmitted_signal2}, we can ensure that $\mathbb{E}[x_k x_j] = 0$ for $k\neq j$ with $k,j\in \mathcal{K}$. Utilizing this fact, the covariance of the received signal is obtained as
	\begin{align}\label{eq:Cov_received}
	\mathbb{E}[{\bf y}{\bf y}^{\sf T}] =\sum_{k\in\tilde{\mathcal{K}}}  \alpha_k{\bf h}_k{\bf h}_k^{\sf T} + \sigma^2{\bf I}_{2M}.
	\end{align}
	In what follows, we characterize ${\bm \Omega}_i$ and then provide a closed-form expression for $\mathbb{E}[\|{\bf r}_i\|^2]$ for three cases discussed in Sec.~\ref{Sec:Proposed}.

	\subsection{Case 1: $\mathcal{S}_i \subset \tilde{\mathcal{K}}$ and $\mathcal{S}_i\neq \tilde{\mathcal{K}}$}
	In this case, ${\bm \Omega}_i$ in \eqref{eq:Omega_def} is expressed as
	\begin{align}\label{eq:Omega_1}
	{\bm \Omega}_i = \left(\sum_{k \in\mathcal{S}_i} \alpha_k {\bf h}_k{\bf h}_k^{\sf T} + \sigma^2{\bf I}_{2M}\right)^{-1}.
	\end{align}
	From \eqref{eq:Omega_1}, the covariance of the received signal in \eqref{eq:Cov_received} is rewritten as
	\begin{align}\label{eq:Cov_received_1}
	\mathbb{E}[{\bf y}{\bf y}^{\sf T}] = {\bm \Omega}_i^{-1}+ \sum_{k \in\tilde{\mathcal{K}}\setminus\mathcal{S}_i} \alpha_k{\bf h}_k{\bf h}_k^{\sf T}.
	\end{align}
	Then the norm squared of the residual vector when the support set belongs to {\bf Case 1} is obtained by applying \eqref{eq:Cov_received_1} into \eqref{eq:Error_r}:
	\begin{align}
	E_1^{(i)} &=\sigma^4\left[{\sf Tr}({\bm \Omega}_i) + 
	\sum_{k \in\tilde{\mathcal{K}}\setminus\mathcal{S}_i} \alpha_k \|{\bm \Omega}_i{\bf h}_k\|^2\right].
	\end{align}

	\subsection{Case 2: $\mathcal{S}_i =\tilde{\mathcal{K}}$}
	In this case, ${\bm \Omega}_i$ in \eqref{eq:Omega_def} is expressed as
	\begin{align}\label{eq:Omega_2}
	{\bm \Omega}_i =\left( \sum_{k \in\tilde{\mathcal{K}}} \alpha_k {\bf h}_k{\bf h}_k^{\sf T} + \sigma^2{\bf I}_{2M}\right)^{-1}.
	\end{align}
	From \eqref{eq:Omega_2}, the covariance of the received signal in \eqref{eq:Cov_received} is rewritten as
	\begin{align}\label{eq:Cov_received_2}
	\mathbb{E}[{\bf y}{\bf y}^{\sf T}] = {\bm \Omega}_i^{-1}.
	\end{align}
	By applying \eqref{eq:Cov_received_2} into \eqref{eq:Error_r}, the norm squared of the residual vector when the support set belongs to {\bf Case 2} is given by 
	\begin{align}
	E_2^{(i)} &=\sigma^4{\sf Tr}({\bm \Omega}_i).
	\end{align}

	\subsection{Case 3: $\mathcal{S}_{i} = \tilde{\mathcal{K}} \cup \{k_i^\star\}$}
	In this case, ${\bm \Omega}_i$ in \eqref{eq:Omega_def} is expressed as
	\begin{align}\label{eq:Omega_3}
	{\bm \Omega}_i =\left( \sum_{k \in\tilde{\mathcal{K}}} \alpha_k {\bf h}_k{\bf h}_k^{\sf T} + \alpha_{k_i^\star} {\bf h}_{k_i^\star}{\bf h}_{k_i^\star}^{\sf T}+ \sigma^2{\bf I}_{2M}\right)^{-1}.
	\end{align}
	From \eqref{eq:Omega_3}, the covariance of the received signal in \eqref{eq:Cov_received} is rewritten as
	\begin{align}\label{eq:Cov_received_3}
	\mathbb{E}[{\bf y}{\bf y}^{\sf T}] = {\bm \Omega}_i^{-1} - \alpha_{k_i^\star} {\bf h}_{k_i^\star}{\bf h}_{k_i^\star}^{\sf T}.
	\end{align}
	Applying \eqref{eq:Cov_received_3} into \eqref{eq:Error_r} yields 
	\begin{align}\label{eq:Cov_e_3_pre}
	E_3^{(i)} &=\sigma^4\left[{\sf Tr}({\bm \Omega}_i) - 
	\alpha_{k_i^\star} \|{\bm \Omega}_i{\bf h}_{k_i^\star} \|^2\right].
	\end{align}  
	From \eqref{eq:Omega_update}, the second term in the RHS of \eqref{eq:Cov_e_3_pre} is expressed as
	\begin{align}\label{eq:Cov_e_3_norm_term} 
	\alpha_{k_i^\star} \|{\bm \Omega}_i{\bf h}_{k_i^\star} \|^2
	= \frac{\alpha_{k_i^\star} \|{\bm \Omega}_{i-1}{\bf h}_{k_i^\star}\|^2}
	{(1 + \alpha_{k_i^\star}{\bf h}_{k_i^\star}^{\sf T} {\bm \Omega}_{i-1}{\bf h}_{k_i^\star})^2}.
	\end{align} 
	Also, applying the trace function to \eqref{eq:Omega_update} yields  
	\begin{align}
	{\sf Tr}({\bm \Omega}_{i}) = {\sf Tr}({\bm \Omega}_{i-1}) 
	-  \frac{\alpha_{k_i^\star}\|{\bm \Omega}_{i-1}{\bf h}_{k_i^\star}\|^2}
	{1 + \alpha_{k_i^\star}{\bf h}_{k_i^\star}^{\sf T} {\bm \Omega}_{i-1}{\bf h}_{k_i^\star}},
	\end{align} 
	so we have
	\begin{align}\label{eq:Cov_e_3_norm_term2} 
	\alpha_{k_i^\star} \|{\bm \Omega}_{i-1}{\bf h}_{k_i^\star} \|^2
	= \frac{{\sf Tr}({\bm \Omega}_{i-1}) - {\sf Tr}({\bm \Omega}_{i}) }
	{1 + \alpha_{k_i^\star}{\bf h}_{k_i^\star}^{\sf T} {\bm \Omega}_{i-1}{\bf h}_{k_i^\star}}.
	\end{align} 
	By applying \eqref{eq:Cov_e_3_norm_term2}  into \eqref{eq:Cov_e_3_pre}, the norm squared of the residual vector when the support set belongs to {\bf Case 3} is given by 
	\begin{align}\label{eq:Cov_y_2}
	E_3^{(i)} &= \sigma^4\left[{\sf Tr}({\bm \Omega}_i) - 
	\frac{{\sf Tr}({\bm \Omega}_{i-1}) - {\sf Tr}({\bm \Omega}_i)}
	{ 1+ \alpha_{k_i^\star}{\bf h}_{k_i^\star}^{\sf T} {\bm \Omega}_{i-1}{\bf h}_{k_i^\star}}\right].
	\end{align}

\end{document}